\documentclass[preprint,12pt]{elsarticle}

\usepackage{graphicx}
\usepackage{caption}
\usepackage{subcaption}
\usepackage{amsmath}
\usepackage{mathalpha}
\usepackage[colorlinks=true,linkcolor=blue]{hyperref}
\usepackage{multirow}
\usepackage{array}
\newcommand{\PreserveBackslash}[1]{\let\temp=\\#1\let\\=\temp}
\newcolumntype{C}[1]{>{\PreserveBackslash\centering}p{#1}}

\usepackage{amssymb}
\usepackage{lineno}

\biboptions{sort&compress}
\journal{---}

\begin{document}

\begin{frontmatter}


\title{Local flame displacement speeds of hydrogen-air premixed flames in moderate to intense turbulence}


\author[fir]{Yuvraj}
\author[sec]{Wonsik Song}
\author[third]{Himanshu Dave}
\author[sec]{Hong G. Im}
\author[fir]{Swetaprovo Chaudhuri \corref{cor4}}
\ead{schaudhuri@utias.utoronto.ca}

\address[fir]{Institute for Aerospace Studies, University of Toronto, Toronto, Canada}
\address[sec]{Clean Combustion Research Center, King Abdullah University of Science and Technology, Thuwal, Saudi Arabia}
\address[third]{Aéro-Thermo-Mécanique, Université libre de Bruxelles, Ixelles, Belgium}
\cortext[cor4]{Corresponding author}

\begin{abstract}
Comprehensive knowledge of local flame displacement speed, $S_d$, in turbulent premixed flames is crucial towards the design and development of hydrogen fuelled next-generation engines. Premixed hydrogen-air flames are characterized by significantly higher laminar flame speed compared to other conventional fuels. Furthermore, in the presence of turbulence, $S_d$ is enhanced much beyond its corresponding unstretched, planar laminar value $S_L$. In this study, the effect of high Karlovitz number ($Ka$) turbulence on density-weighted flame displacement speed, $\widetilde{S_d}$, in a H$_2$-air flame is investigated. Recently, it has been identified that flame-flame interactions in regions of large negative curvature govern large deviations of $\widetilde{S_d}$ from $S_L$, for moderately turbulent flames. An interaction model for the same has also been proposed. In this work, we seek to test the interaction model's applicability to intensely turbulent flames characterized by large $Ka$. To that end, we investigate the local flame structures: thermal, chemical structure, the effect of curvature, along the direction that is normal to the chosen isothermal surfaces. Furthermore, relative contributions of the transport and chemistry terms to $\widetilde{S_d}$ are also analyzed. It is found that, unlike the moderately turbulent premixed flames, where enhanced $\widetilde{S_d}$ is driven by interactions among complete flame structures, $\widetilde{S_d}$ enhancement in high $Re_t$ and high $Ka$ flame is predominantly governed by local interactions of the isotherms. It is found that enhancement in $\widetilde{S_d}$ in regions of large negative curvature occurs as a result of these interactions, evincing that the interaction model is useful for high $Ka$ turbulent premixed flames as well.

\end{abstract}

\begin{keyword}
flame displacement speed \sep turbulent premixed flames \sep interacting flames


\end{keyword}

\end{frontmatter}

%\linenumbers

\section{Introduction}
\label{S:1}
Climate change is undoubtedly the greatest challenge faced by humanity. Aircraft engines contribute to about 3.5\% of total radiative forcing by all human activities \cite{penner1999}. However, unlike other greenhouse gas emitters, there has been no real alternative to the easily accessible, energy-dense, liquid hydrocarbon fuels for long-range aircraft propulsion. Recently, hydrogen is being strongly considered as a potential gas turbine fuel for long-haul aircraft due to its unmatched energy density and zero greenhouse gas emissions. One of the significant challenges associated with hydrogen, apart from storage and delivery, is its fast unstretched, planar, laminar flame speed, $S_L$. Gas turbine combustors have intense turbulent flow fields with the combustor turbulence Reynolds number, $Re_t \approx 50000-100000$. In addition to the already large flame speed, turbulence is certain to further enhance the local flame speeds of hydrogen-air premixed flames through stretching, folding, collisions or turbulent diffusion. Such flame speed enhancements can exacerbate the potential problems of hydrogen-air combustion in gas turbines due to phenomena such as flashback, thermo-acoustic instability, to name a few. Thus, efforts to develop hydrogen-based combustors require investigation of several fundamentals combustion topics such as in-depth knowledge of flame speeds of hydrogen-air premixed flames in turbulence, understanding the causality of its enhancement, corresponding limits, and developing predictive physics-based models. 

Local flame displacement speed, $S_d$, is the speed at which a point on a flame surface moves along the local normal relative to the local flow velocity. Stretching, folding, and collision of flame surfaces on interaction with turbulent eddies have a considerable effect on $S_d$. Over the years, dependence of $S_d$ on curvature \cite{echekki1996, echekki1999, im2000effects, chen1998correlation, cifuentes2018, chakraborty2005_1, uranakara2016, uranakara2017}, strain \cite{echekki1996, chakraborty2005_1} and the Lewis number $Le$ \cite{haworth1992, rutland1993, chakraborty2005_2, alqallaf2019} for moderately turbulent premixed flames have been studied extensively.
Recently, \citet{yu2021components} presented a new set of transport and surface averaged evolution equations for the different terms in the decomposition of the displacement speed, $S_d$, of an iso-scalar surface in a turbulent reacting flow. Low and moderate $Ka$ DNS datasets were examined using newly derived averaged equations. Their findings indicated that the key terms in the considered evolution equations were weakly affected by thermal expansion. Furthermore, apart from introducing a dilation term, the thermal expansion also caused variations in surface averaged flame speed through curvature/dilatation correlation and strain-rate/curvature-tensor correlation. Unlike the other equations, the terms in the curvature equation remained unaffected by the flow. \citet{attili2021} and \citet{wang2017} investigated flame structures for premixed jet flames that eventually led to the enhancement of $S_d$ at moderate and high $Ka$. A number of ``flame-in-the-box" simulations by \citet{hamlington2011, poludnenko2010, poludnenko2011,song2020dns,lipatnikov2021}
explored a wide range of $Ka$ conditions reaching up to thin reaction zone and distributed regimes, and reported that even the strong turbulence broadened the preheat zone only while the main reaction zones remained intact. It was found that the global acceleration of the flame was obtained as a combined effect of an increase in the flame surface area along with flame-flame collision and large flame curvature \cite{poludnenko2011,song2020dns}. For detailed insights into preheat zone structure at high $Ka$ turbulence, the readers could refer to the recent review by \citet{driscoll2020}.

The asymptotic analysis \cite{pelce1982, matalon1982,matalon1983,candel1990} and experimental measurements \cite{wu1985, egolfopoulos1989, law2000} established the explicit relation between $S_d$ and the flame stretch, $\mathbb{K}$, which consists of the strain rate, $K_s$, and the contribution of the curvature, $\kappa$, for laminar premixed flames. The linearized theory at low stretch conditions yielded \cite{matalon1982,matalon1983}: 
% Add reference Matalon (1983) "On Flame Stretch"
%induced by the velocity field. The combined effect of both the parameters was taken into account in the form of stretch-rate $\mathbb{K}$, which formed a linear relationship with $S_d$ given by Eq.~\ref{eq: Sd_eq_1}. Here, the constant $\mathcal{L}$ is the Markstein Length.
%----eq 1-----
\begin{equation}\label{eq: Sd_eq_1}
    S_{d,u} = S_L - \mathcal{L}\mathbb{K} %\kappa
\end{equation}
where $S_{d,u}$ is $S_d$ conditioned on the unburnt side of the flame, and 
\begin{equation}\label{eq: curv}
\mathbb{K} = K_s + S_L \kappa
\end{equation}
and $\kappa = \nabla \cdot \boldsymbol{n}$ is the curvature. The proportionality constant, $\mathcal{L}$, was referred to as the Markstein length \cite{markstein1964} which is a function of the mixture properties.
Since $S_d$ is defined at an iso-surface of a scalar variable (e.g. temperature), the dependence of $S_d$ on the local density is minimized by defining the density-weighted displacement speed, $\widetilde{S_d} =\rho_0 S_d/\rho_u$, where $\rho_0$ is the density at the isotherm and $\rho_u$ is the density of fresh reactants \cite{echekki1996,im1999}.

Recognizing that the flame responds differently to the tangential strain rate and curvature, more recent studies, experimental work by \citet{bradley1996}, theoretical work by \citet{bechtold2001} and \citet{clavin2011} distinguished two different Markstein lengths to represent the effects of ${K_s}$ and $\kappa$. In particular, \citet{bechtold2001} derived:
%------eq 2----

\begin{equation} \label{eq: Sd_eq_2}
    S_{d,u} = S_L - \mathcal{L}_{s}K_s - \mathcal{L}_\kappa S_L\kappa
\end{equation}

Recently, \citet{giannakopoulos2015_1, giannakopoulos2015_2} derived the following expression for $\widetilde{S_d}(\theta^*)$ conditioned on any isotherm inside the flame

%-------eq 3----------
\begin{equation} \label{eq: Sd_eq_3}
    \widetilde{S_d}(\theta^*) = S_L - \mathcal{L}_{\mathbb{K}}(\theta^*)\mathbb{K} - \mathcal{L}_\kappa(\theta^*)S_L\kappa
\end{equation}

where $\theta^* = T_0/T_u$ is the non-dimensional temperature, where $T_0$ is the temperature at the iso-surface and $T_u$ is the temperature of the unburnt reactants. The $\mathcal{L}_{s}$, $\mathcal{L}_{\mathbb{K}}$, and $\mathcal{L}_\kappa$ are the Markstein length for strain rate, total stretch rate and curvature, respectively, calculated from theoretical expressions, in the linear, weak stretch regime \cite{giannakopoulos2015_2}. 

Using the Lagrangian flame particles as a representative of local states of a flame, \citet{dave2020} employed the forward \cite{chaudhuri2015} and backward \cite{dave2018} flame particle tracking (FPT) methods to examine the evolution of specific flame front points from their inception to annihilation for moderately turbulent premixed ($Ka \sim \mathcal{O}(10)$) H$_2$-air flames. The FPT analysis applied to a large number of particles allowed the identification of a manifold of states that represent the statistical characteristics of the turbulent flame segments for a given inlet flow and thermodynamic conditions. The study found that, while the two Markstein length, weak stretch flame speed model was valid for the majority of particles over a significant part of their lifetime, it failed to capture the large variation in $S_d$ towards the end of its lifetime, leading to annihilation. These variations were found to be a result of flame-flame interactions in regions of large negative curvature \cite{dave2020}. Subsequently, they investigated these self-interactions by analyzing the transient structure of a collapsing laminar cylindrical flame with $Le=1$, as a representative of the local structure of the interacting flame elements in turbulence. After invoking the constant property assumption, the functional form of the $S_d$ relation during flame-flame interaction with negative $\kappa$, was obtained as:

%-----eq4---
\begin{equation}\label{eq: Sd_eq_I}
    S_d = -2\alpha_u \kappa
\end{equation}

Here $\alpha_u$ is the thermal diffusivity of the mixture at inlet conditions. Using $\alpha_0$: the isotherm ($T=T_0$) conditioned thermal diffusivity instead of $\alpha_u$, the leading order $S_d$ expression from Eq. \ref{eq: Sd_eq_I} can be rewritten as: 
\begin{equation}\label{eq: Sd_eq_I0}
    \frac{S_d \rho_0}{\rho_u} = -2\alpha_0 \frac{\rho_0}{\rho_u}  \kappa = -2\widetilde{\alpha_0}\kappa
\end{equation}
However, this equation is valid only for $\kappa<<0$. Assuming that to the leading order, as $\kappa \rightarrow 0$, $\widetilde{S_d} \rightarrow S_L$, we can write
\begin{equation}\label{eq: Sd_eq_I00}
    \widetilde{S_d} = S_L -2\widetilde{\alpha_0} \kappa 
\end{equation}
$\widetilde{\alpha_0}$ is the density weighted thermal diffusivity conditioned on $T_0$. Alternatively, starting from Eq. \ref{eq: Sd_eq_I} and applying scaling analysis of the RHS of the energy equation (Eq. \ref{eq: Sd_eq_DNS_T}), the following interaction model for $\widetilde{S_d}$ was proposed in \cite{dave2020}: 

%---eq 4----
\begin{equation}\label{eq: Sd_eq_I2}
    \widetilde{S_d} = S_L - \alpha_0(1+C)\kappa\frac{T_u}{T_0}
\end{equation}

$C$ is a constant. For moderately turbulent flames, the weak stretch rate model with two Markstein lengths (Eq. \ref{eq: Sd_eq_3}) was able to describe large portions of the flame that were free from flame-flame interactions and characterized by $\widetilde{S_d}/S_L$ in the neighborhood of unity \cite{dave2020}. In the interacting, annihilating locations, however, the instantaneous flame structure did not resemble that of a typical stretched flame and hence the weak stretch rate theory was no longer applicable. The interacting flame theory and the corresponding local flame speed given by Eq. \ref{eq: Sd_eq_I2} was found to capture the flame-flame interaction behavior very well for moderately turbulent premixed flames ($Ka \sim \mathcal{O}(10)$). However, its applicability to intense turbulent flames (($Ka\sim \mathcal{O}$(100) and up) is yet to be examined. An $S_d$ model as a function of curvature was proposed by Peters \cite{Peters2000} in the thin reaction zone regime. Recently, using theoretical arguments \citet{trivedi2019topology} suggested that $S_d$ is functionally dependent on the inverse of the local radius of curvature during a tunnel closure event. In both cases, the right-hand side terms of the flame displacement speed equation were computed based on a constant species mass fraction surface.

In view of the above developments, the present study undertakes the detailed analysis using a recent DNS dataset \cite{song2020dns} covering a large range of $Ka$ and $Re_t$, in order to assess the validity of the interaction model at high $Ka\sim \mathcal{O}(1000)$ conditions. The DNS data correspond to different turbulent combustion regimes for a H$_2$-air premixed flame: two in broken reaction regime and two in thin reaction regime. We first present the joint probability density functions (JPDFs) of $S_d$ and $\kappa$ from DNS alongside its comparison with the interaction model. This is followed by analysis of the local chemical and thermal structure and the nature of flame-flame interaction in different regimes. The relative contributions of the transport and chemistry parameters enhancing local $S_d$ are also examined. This is accomplished by investigating the different terms in Eq. \ref{eq: Sd_eq_DNS_T} for $S_d$, such as the heat conduction, heat transport via species diffusion, and the heat release rate.

% WSONG
\section{Computational Methodology}
\label{S:2}

\begin{table}[h!]
\footnotesize
 \begin{center}
 \begin{tabular}{ | p{4cm} | p{1.8cm} | p{1.8cm} | p{1.8cm} | p{1.8cm} |} 
 \hline
 Parameters & F1 & F2 & F3 & F4 \\ 
  \hline
  \hline
 Domain dimensions [cm] & $2\times1\times1$ & 0.427 & 0.918 & 0.338 \\
  & & $\times0.146$ & $\times0.368$ & $\times0.052$\\
  & & $\times0.146$ & $\times0.368$ & $\times0.052$\\
 Grid points & $1000\times500$ & $1645\times560$ & $460\times184$ & $1300\times200$ \\
  & $\times500$ & $\times560$  & $\times184$ & $\times200$ \\
 Integral length scale, $l_o$ [cm] & 0.200 & 0.029 & 0.074 & 0.010 \\
 Root mean square velocity, $u_{rms}$ [cm/s] & 678.1 & 4746.7 & 359.1   & 2481.8 \\
 Kolmogorov length scale, $\eta$ [$\mu$m] & 14.916 & 2.141 & 18.714 & 2.683 \\
 Karlovitz no., $Ka$ &  23.2 &  1127.3 &  14.7 & 716.7 \\
 Reynolds no., $Re_t$& 686.6 & 697.5 & 133.9 & 128.9 \\
 Damk\" ohler no., $Da$ & 1.130 & 0.023 & 0.785 & 0.016 \\
 $\Delta x/\eta$ & 1.341 & 1.214 & 1.069 & 0.969 \\
 $\delta_L/\Delta x$ & 17.7 & 136.2 & 17.7 & 136.2 \\
 $\delta t$ [$\mu$s] & 1E-02 & 1E-03    & 5E-03 & 5E-4\\
 \hline
 \end{tabular}
 \caption{\label{tab:1}Details of the parameters for the four DNS cases studied in this paper. For all the cases, $T_u$ = 300 K, $P$ = 1 atm, $\phi$ = 0.7, $\delta_L=(T_b - T_u)/|\nabla T|_{max}$ = 0.03541 cm, $S_L$ = 135.619 cm/s}
 \end{center}
 \end{table}

\begin{figure}[h!]
\centering\includegraphics[trim=2.5cm 9cm 2cm 8.1cm,clip,width=1.0\textwidth]{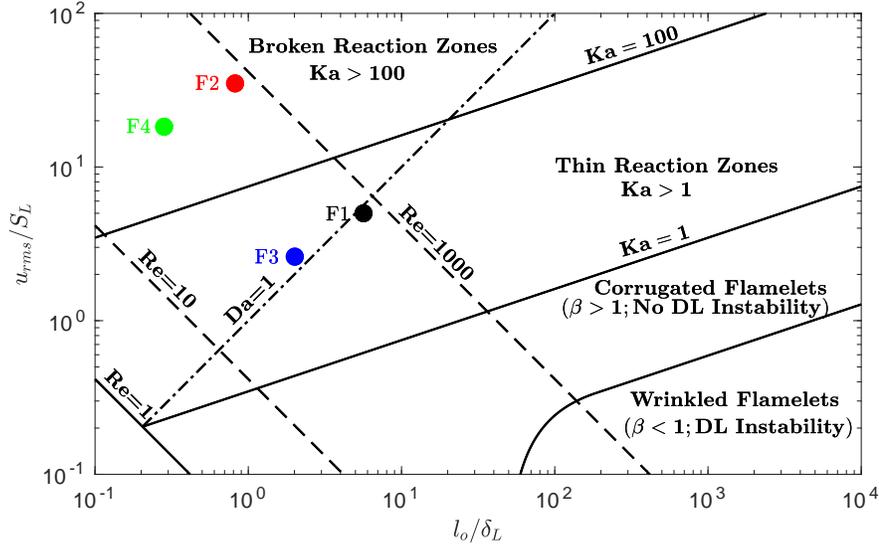}
\caption{Regime diagram specifying the cases: F1, F2, F3, F4}
\label{Fig:regime_diagram}
\end{figure}

% \subsection{Direct Numerical Simulation}
The DNS data listed and shown in Table~\ref{tab:1} and Fig.~\ref{Fig:regime_diagram}, respectively, are generated in a previous study by \citet{song2020dns}. A series of simulations with lean H$_2$-air premixed flames propagating into forced turbulence are performed in a turbulence-in-a-box configuration by changing the turbulence parameters. Different combinations of integral length scale $l_o$ and root-mean-square (RMS) turbulent velocity fluctuations $u_{rms}$ were used to validate the existing theory of modeling flame displacement speed at highly turbulent conditions. The resulting turbulent Reynolds number ($Re_t$) and Karlovitz number ($Ka$) vary in orders of magnitude; $Re_t = u_{rms}l_0/\nu$, where $\nu$ is the kinematic viscosity of the reactant mixture  while $Ka = \tau_f/\tau_{\eta}$ is the ratio of the flame time scale $\tau_f = \delta_L/S_L$  to the Kolmogorov time scale $\tau_{\eta}$ in the unburned mixture. Here, $\delta_L=(T_b - T_u)/|\nabla T|_{max}$ is the thermal thickness of the corresponding standard premixed flame, $T_b$ being the burnt gas temperature. The cases F1 to F4 correspond to a cuboid configuration. The cases F1 and F3 are located in the thin reaction zone regime, while the cases F2 and F4 are located in the broken reaction zone regime in the modified Borghi-Peters turbulent regime diagram \cite{chaudhuri2011} shown in Fig. \ref{Fig:regime_diagram}. The cases F1 and F2 have nearly the same $Re_t \approx 700$ while the cases F3 and F4 have $Re_t \approx  130$, respectively, but have significantly different $Ka$. The cases F1 and F3 have $10<Ka<20$, whereas the cases F2 and F4 have $700<Ka<1200$.

The conservation of mass, momentum, energy and species equations are solved with a fully compressible solver called KAUST Adaptive Reacting Flows Solver (KARFS)~\cite{perez2018direct}. An eighth-order central-difference scheme and a fourth-order Runge-Kutta scheme are applied to the spatial discretization and time integration, respectively. Moreover, a tenth-order filter is used to remove spurious high-frequency fluctuations. Finally, the non-reflecting Navier-Stokes characteristic boundary conditions (NSCBC)~\cite{yoo2005characteristic,yoo2007characteristic} are applied to the outflow boundary, while periodic boundary conditions are applied to the transverse directions.

The turbulent combustion simulations start with the precomputed one-dimensional premixed flame solution using the PREMIX solver~\cite{kee1985premix} in the Chemkin package~\cite{kee1996chemkin} for the same mixture and thermodynamic conditions as the turbulent flames ($\phi$ = 0.7, $T_u$ = 300 K, and $P$ = 1 atm), and the turbulent velocity fluctuations are superimposed. Initial homogeneous and isotropic turbulence field is also generated in a separate simulation in the spectral space by specifying the energy spectrum \cite{passot1987numerical}. The turbulent forcing scheme introduced by \cite{bassenne2016constant} is utilized to maintain the initial turbulence level. The forcing is only activated from 10\% in the $x$-direction of the computational domain and immediately turned off at the flame base defined as $T_\mathrm{cut}$ = 320 K to avoid artificial modification of the flame front by the thermal expansion. The detailed chemical kinetics model, consisting of 9 species and 23 reactions by~\citet{burke2012comprehensive}  is used.

With the direct numerical simulation (DNS) data, $S_d$ is computed locally on the iso-scalar surface, based on the mass fraction of the $k$th species, $Y_k$, or temperature, $T$ as the scalar variable \cite{poinsot2005}. Temperature is often considered a more fundamental quantity when compared to the species mass fraction in the determination of flame speed, as evident from the Zeldovich Frank-Kamenetskii thermal flame theory. Thus, in the present study, the temperature was chosen as the scalar variable, and $S_d$ was determined by the following equation:

\begin{equation}
\begin{split}
    S_{d,T} & = \frac{1}{\rho C_p |\nabla T|}\left[ \nabla\cdot(\lambda^\prime\nabla T) + \rho \nabla T \cdot \sum_{k}^{}  (\mathcal{D}_k C_{p,k} \nabla Y_k) - \sum_{k}^{} h_k \dot{\omega}_k\right] \\
    & = \frac{1}{\rho C_p }\left[ \underbrace{\frac{1}{|\nabla T|}\nabla\cdot(\lambda^\prime\nabla T)}_\text{$T_1$} 
    +  \underbrace{\dfrac{\rho\nabla T}{|\nabla T|} \cdot \sum_{k}^{}  (\mathcal{D}_k C_{p,k} \nabla Y_k)}_\text{$T_2$} 
    - \underbrace{\frac{1}{|\nabla T|}\sum_{k}^{} h_k \dot{\omega}_k}_\text{$T_3$}\right] 
    \end{split}
    \label{eq: Sd_eq_DNS_T}
\end{equation}
where $C_{p,k}$ and $C_{p}$ are the constant-pressure specific heat for the $k$th species and the bulk mixture, respectively, where the mixture-averaged formula was used; $\lambda^\prime$ is the thermal conductivity of the mixture; $\mathcal{D}_k$, $h_k$ and $\dot{\omega}_k$ are the mass diffusivity, enthalpy, and net production rate of the $k$th species, respectively, where local unit normal vector ($\boldsymbol{n}$) 
\begin{equation} \label{eq:normal}
    \boldsymbol{n} = -\frac{\nabla T}{|\nabla T|}
\end{equation}
is defined positive in the direction of the reactant mixture.

%Dependence of $S_d$ on curvature for the overall flame, in the thin reaction regime, was proposed by Peters \cite{Peters2000}. Recently, using theoretical arguments \citet{trivedi2019topology} suggested that $S_d$ is functionally dependent on the inverse of the local radius of curvature during a tunnel closure event. In both cases, the right hand side terms of the flame displacement speed equation were computed based on a constant species mass fraction surface as shown in Eq.~\ref{eq: Sd_eq_DNS_Y}.

\section{Results and Discussion}

\label{S:4}

\subsection{Local flame speed and structure}

\begin{figure}[h!]
\centering\includegraphics[trim=2.0cm 7.5cm 2.0cm 2.5cm,clip,width=1.0\textwidth]{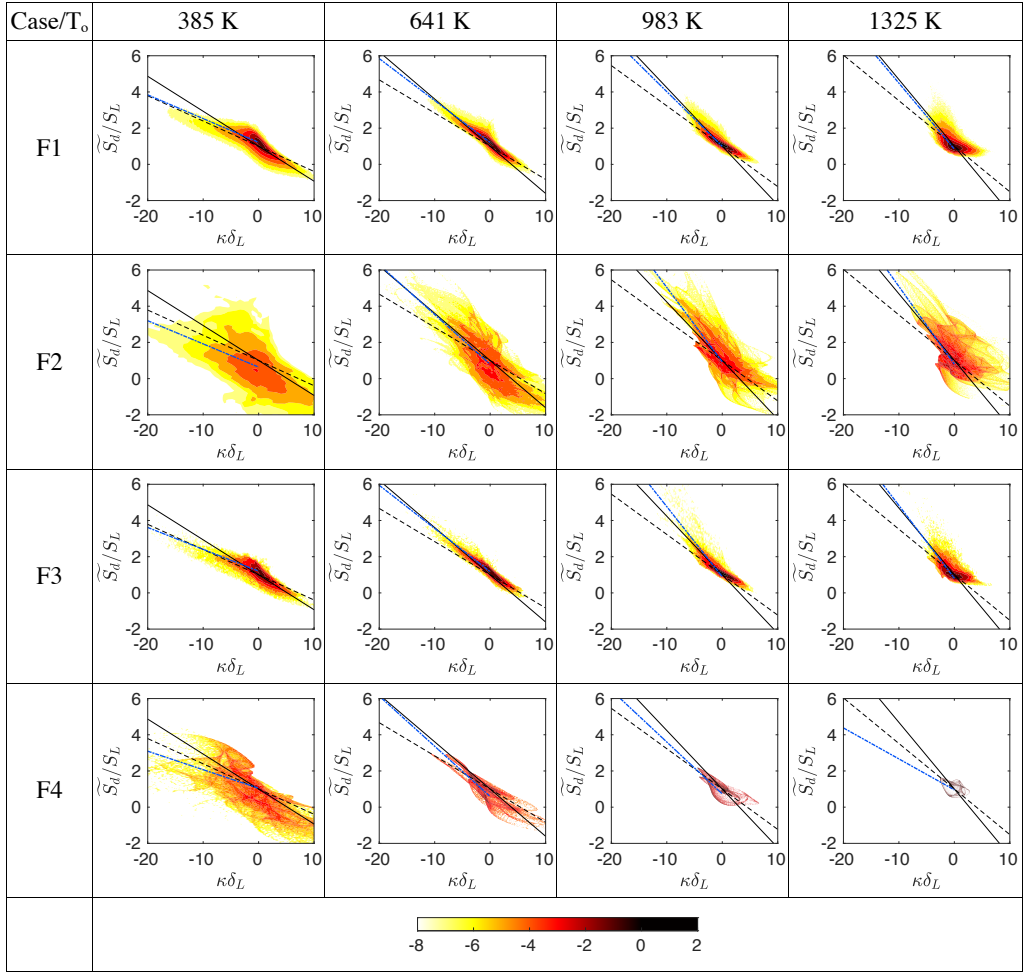}
\caption{Joint probability density function (JPDF) of normalized density weighted flame displacement speed, $\widetilde{S_d}/S_L$ and normalized curvature $\kappa \delta_L$. Colorscale represents natural logarithm of JPDF magnitudes. The solid black line represents the analytical result given by Eq. \ref{eq: Sd_eq_I00}. The dashed black line represents the analytical expression combined with scaling analysis given by Eq. \ref{eq: Sd_eq_I2}, where $C $ = 0.5. Both expressions obtained from cylindrical laminar flame colliding onto itself. The chain dotted blue line is the linear fit of the data points with $\kappa\delta_L<0$.}
\label{Fig:SdSL_kappa}
\end{figure}

\begin{figure}[h!]
    \centering
    \includegraphics[trim=1.cm 0cm 0cm 0cm]{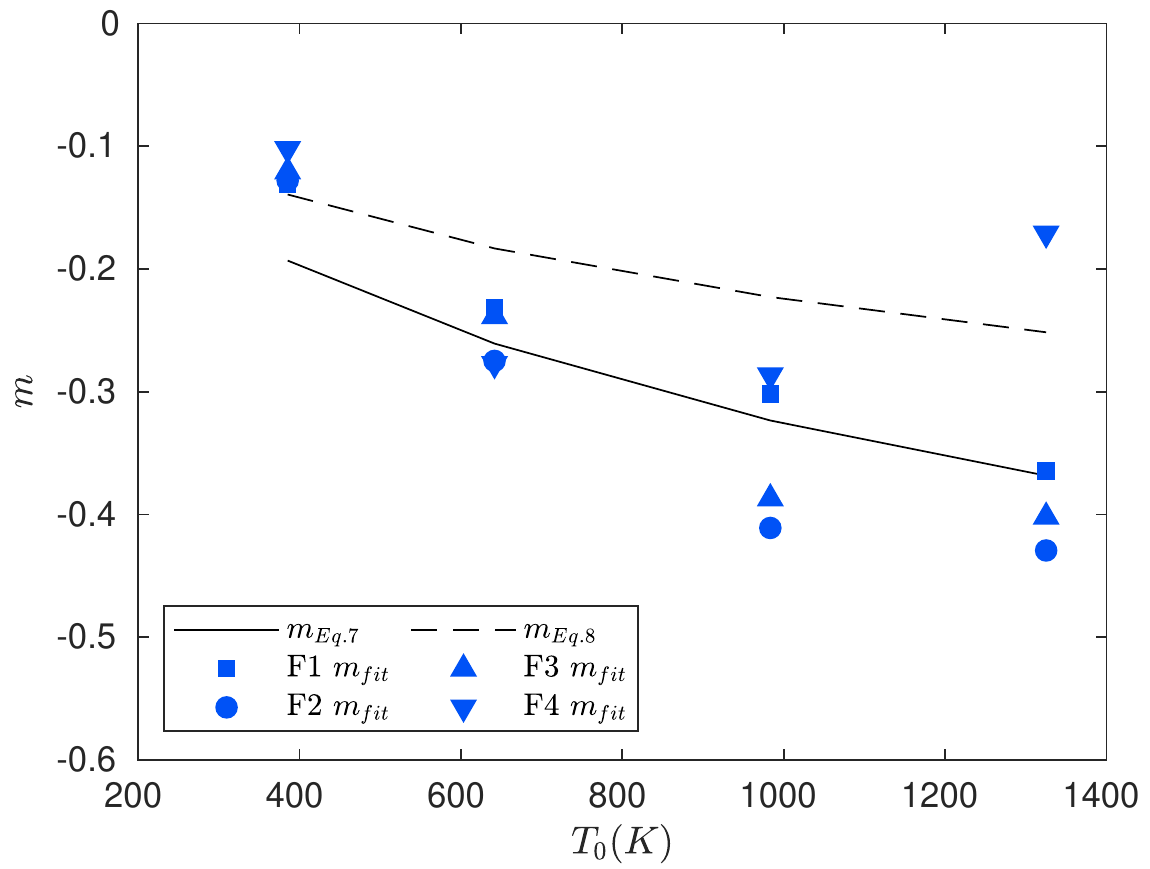}
    \caption{Slopes ($m$) of the line obtained from analytical result by Eq.~\ref{eq: Sd_eq_I00} (solid black line), Eq.~\ref{eq: Sd_eq_I2} with $C $ = 0.5 (dashed black line), and from the line of best fit (blue symbols) vs level set temperature $T_0$ from Fig.~\ref{Fig:SdSL_kappa} for Cases F1-F4.}
    \label{fig:Slope_comparison_theory_fit}
\end{figure}

In this section, we explore the results concerning the enhancement of density-weighted local flame displacement speed, $\widetilde{S_d}$, in turbulent flames (cases F1-F4), over its nominal value given by the flame speed of a standard premixed flame, $S_L$. A standard premixed flame is a one-dimensional, unstretched, laminar premixed flame in a doubly infinite domain~\cite{law2006}. The joint probability density function (JPDF) in terms of $\widetilde{S_d} / S_L$ and non-dimensional curvature, $\kappa\delta_L$, computed for all local points lying on a given flame surface for four different isotherms ($T_0$) ranging from 385 K to 1325 K is shown in Fig.~\ref{Fig:SdSL_kappa}. The color bar represents the natural logarithm of the joint probability. For cases F1-F4, $Re_t$ ranges from 127 to 700 and $Ka$ from 14 to 1126. Overall it is clearly observed that $\widetilde{S_d} / S_L$ is negatively correlated with $\kappa\delta_L$ for all conditions, at all choices of the isotherms. Indeed larger scatters are found for higher $Ka$ conditions for Cases F4 (larger) and F2 (the largest $Ka$). Each of the sixteen sub-figures of Fig.~\ref{Fig:SdSL_kappa} also includes i) a solid black line representing the analytical result given by Eq. \ref{eq: Sd_eq_I00}. ii) A dashed black line representing the analytical expression combined with scaling analysis given by Eq. \ref{eq: Sd_eq_I2}. iii) The linear fit of the $\widetilde{S_d} / S_L - \kappa\delta_L$ DNS data for $\kappa\delta_L <$ 0 shown by a chain-dotted blue line. Each solid black line represents Eq. \ref{eq: Sd_eq_I00} and the dashed black line represents Eq. \ref{eq: Sd_eq_I2}, both obtained from the interacting flame theory, and are superposed on each individual JPDF. It is evident that both theoretical expressions capture the overall trends of the DNS data closely, showing good agreement with the corresponding lines of best fit. 
The slopes of the theoretical lines given by $m_{Eq. \ref{eq: Sd_eq_I00}}=-2\widetilde{\alpha_0}/(S_L\delta_L)$ from Eq. \ref{eq: Sd_eq_I00} and $m_{Eq. \ref{eq: Sd_eq_I2}}=-\alpha_0(1+C)T_u/(T_0 S_L\delta_L)$ with $C=0.5$ from Eq. \ref{eq: Sd_eq_I2}, respectively, are compared with the slopes of lines of the best fit ($m_{fit}$) for all the cases at different $T_0$ in Fig. \ref{fig:Slope_comparison_theory_fit}. Given the complexity embedded in $S_d$ in intensely turbulent flows, the $m_{Eq. \ref{eq: Sd_eq_I00}}$ and $m_{Eq. \ref{eq: Sd_eq_I2}}$ both capture $m_{fit}$ and its trends reasonably well. In particular, the agreement between $m_{Eq. \ref{eq: Sd_eq_I2}}$ and $m_{fit}$ is excellent at smaller values of $T_0$ and worsens with increasing $T_0$. Overall, the agreement between $m_{Eq.\ref{eq: Sd_eq_I00}}$ and $m_{fit}$ is better at most $T_0$ values. Some of the reasons for the deviations could be attributed to near planar collisions, which can enhance $S_d$ even at small $\kappa$. Furthermore, the theory does not account for the interaction of the heat release layers. This point will be taken up in later discussions.  %The $m_{Eq. \ref{eq: Sd_eq_I2}}$ for a given $T_0$ shows small variation across the different cases. This is due to the small differences in averaged thermal diffusivity conditioned on the respective isotherms.   
%The theory given Eq.~\ref{eq: Sd_eq_I2} is in close agreement with the best fit lines for most isotherms from all the cases F1-F4 - a total of 15 conditions, except for the F4 case with $T_0=1325K$. Here, the lack of data points over the extended ranges of $S_d$ and $\kappa$, as available in other cases, did not yield a meaningful fit. This is due to fast turbulence dissipation at high $Ka$, low $Re_t$ and will be revisited later.
Qualitatively, Eq. \ref{eq: Sd_eq_I00} and \ref{eq: Sd_eq_I2} correctly capture the decreasing slope with an increase in $T_0$, evident from the $m_{fit}$ data points. This is attributed to the increasing density weighted thermal diffusivity of the mixture at higher temperatures. The only exception is the $m_{fit}$ for F4 case with $T_0=1325K$. Here, the lack of data points over the extended ranges of $S_d$ and $\kappa$, as available in other cases, probably did not yield a meaningful fit. This is due to fast turbulence dissipation at high $Ka$, low $Re_t$ and will be revisited later. $\alpha_0$ at different $T_0$ are obtained from the Chemkin PREMIX calculation of the corresponding 1-D standard premixed flame \cite{kee1985premix}. 
Figure~\ref{Fig:Snapshots_flame_Sd_kappa} provides visual evidence of increased enhancement of $\widetilde{S_d}/S_L$ at large negative curvature, for Cases F1-F4, as seen in the JPDFs.

\begin{figure}[h!]
\begin{subfigure}[b]{1.0\textwidth}
\centering\includegraphics[trim=0cm 4.cm 0cm 0.9cm,clip,width=1.0\textwidth]{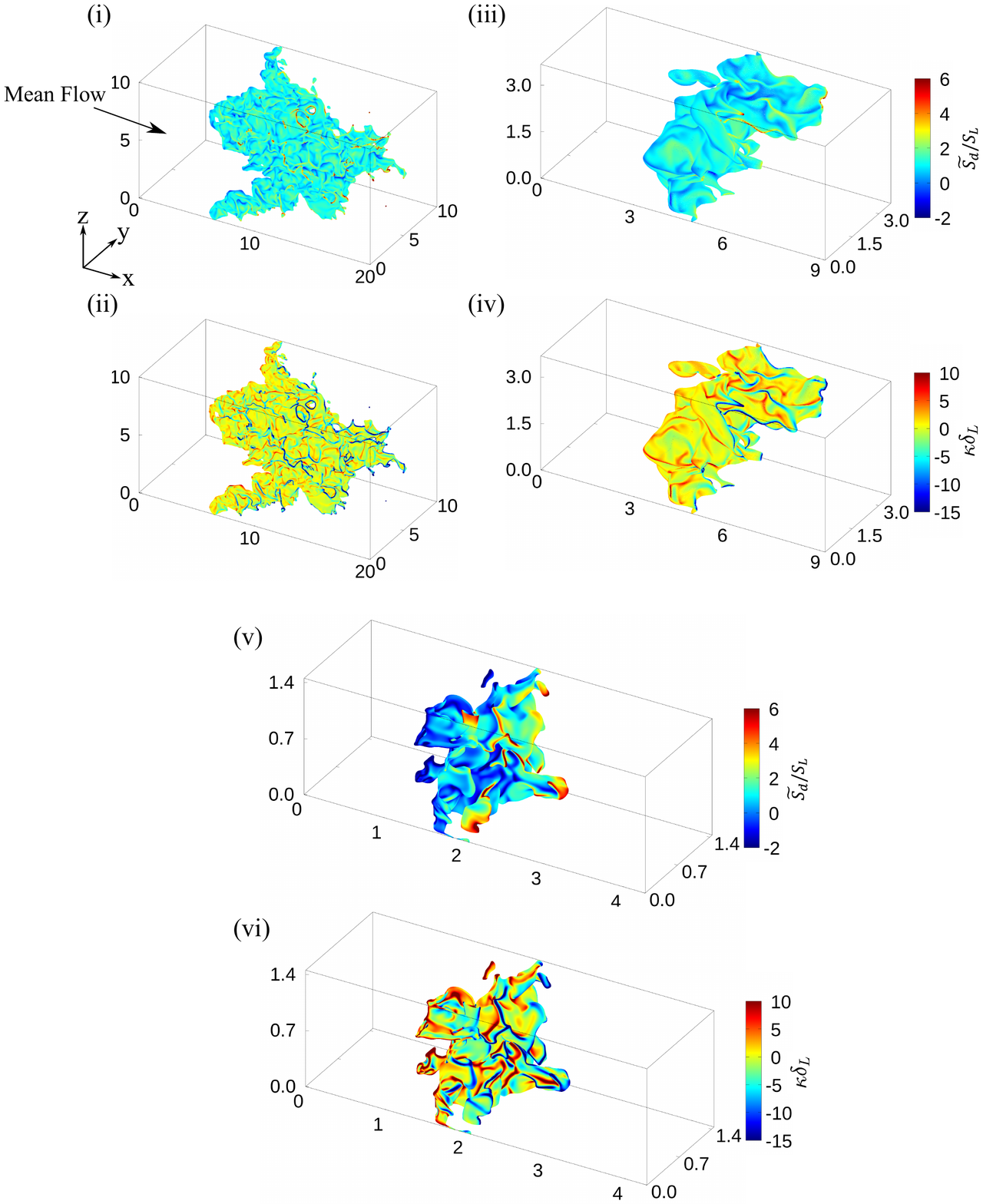}
\caption{Flame surface represented by $T_0 $ = 385 K, for F1 (i, ii), F3 (iii, iv) and F2 (v, vi) colored by (i, iii, v) $\widetilde{S_d}/S_L$ (ii, iv, vi) $\kappa\delta_L$ }
\end{subfigure}
\end{figure}

\begin{figure}[h!]\ContinuedFloat
\begin{subfigure}[b]{1.0\textwidth}
\centering\includegraphics[trim=0.4cm 9cm 0cm 1cm,clip,width=1.0\textwidth]{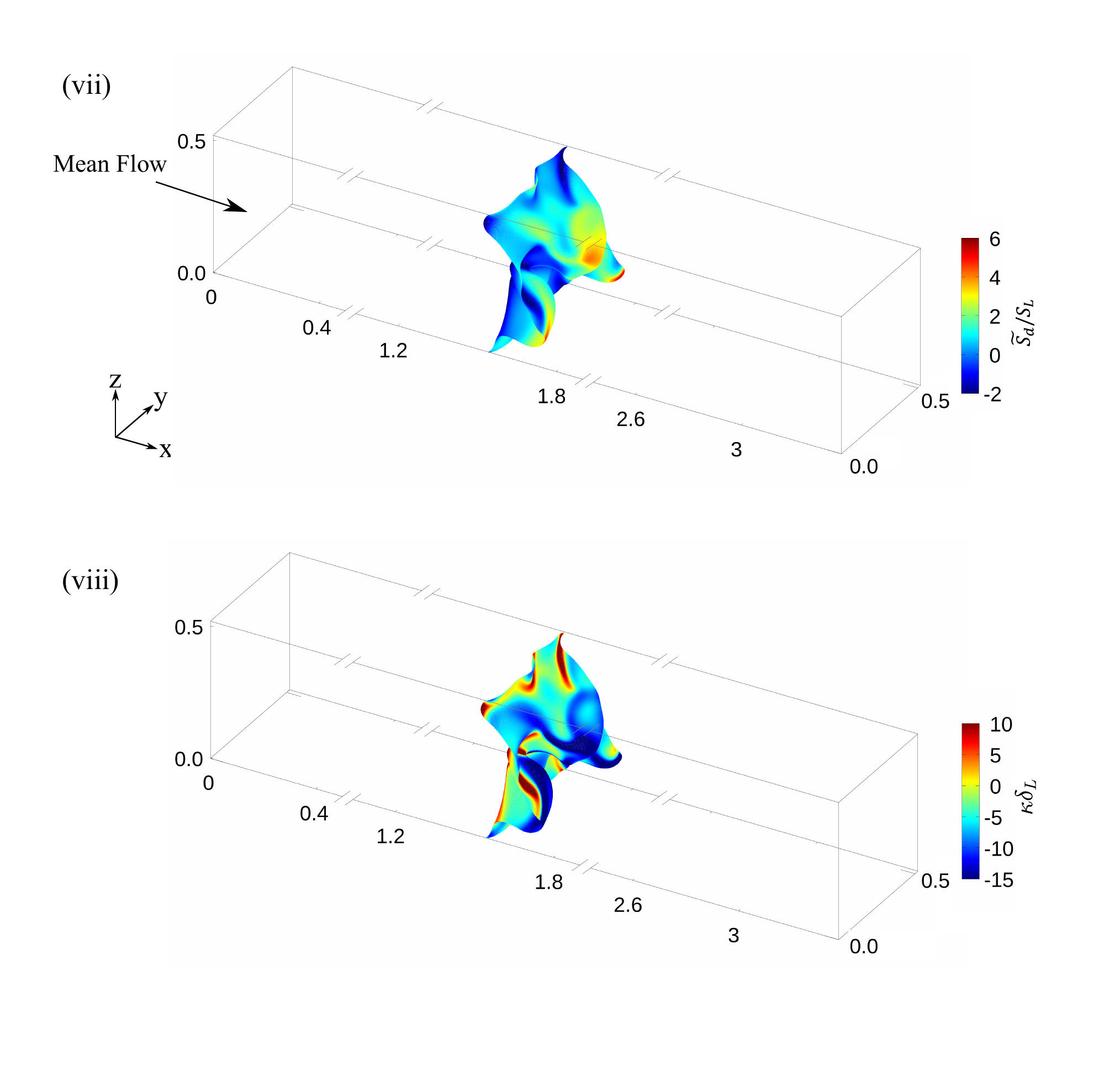}
\caption{Flame surface represented by $T_0$ = 385 K, for F4  colored with (vii) $\widetilde{S_d}/S_L$ (viii) $\kappa\delta_L$ }
\end{subfigure}
\caption{Instantaneous snapshot of the turbulent flame(represented by $T_0 = $385 K) for F4. The flame surfaces are colored with (vii) $\widetilde{S_d}/S_L$ and (viii) $\kappa\delta_L$ to mark the locations on the flame surface possessing large curvature and local flame speed.}
\label{Fig:Snapshots_flame_Sd_kappa}
\end{figure}

Despite the good agreement between the predictions by the interacting flame theory and the DNS results, the generality of the theory needs to be further scrutinized. %Combining the two-Markstein length flame speed model with an interacting flame speed model, an integrated theoretical model \cite{dave2020} has been shown to predict the flame displacement speeds for moderately turbulent premixed flames. 
The interacting flame speed model was derived assuming a laminar, cylindrical premixed flame configuration that propagates inwardly to eventually interact with itself and annihilate. On the other hand, the DNS results under study are obtained over a very large range of $Re_t$ and $Ka$ representing intensely turbulent conditions, where the local flame structure is not necessarily in the form of a self-annihilating cylindrical laminar flame. Thus, a question arises as to whether the reasonably accurate predictions by the theory are merely fortuitous or if the theory properly captures the essential underlying physics responsible for the enhanced $\widetilde{S_d}$ in the most general sense, even in high $Ka$ turbulent flames.

To answer this question, the local flame structure is examined for all the cases in order to address the following specific points:
\begin{enumerate}
    \item Is the enhancement of $\widetilde{S_d}/S_L$ in negatively curved flame regions driven by the flame-flame interactions/collisions? If so, the DNS data must reveal statistical evidence that the corresponding regions exhibit the thermo-chemical structure of interacting flames.
    \item Does the local turbulent flame structure, interacting or non-interacting, resemble that of a corresponding laminar flame?
\end{enumerate}

\begin{figure}[h!]
\centering\includegraphics[trim=2cm 17.5cm 1.5cm 1.5cm,clip,width=1\textwidth]{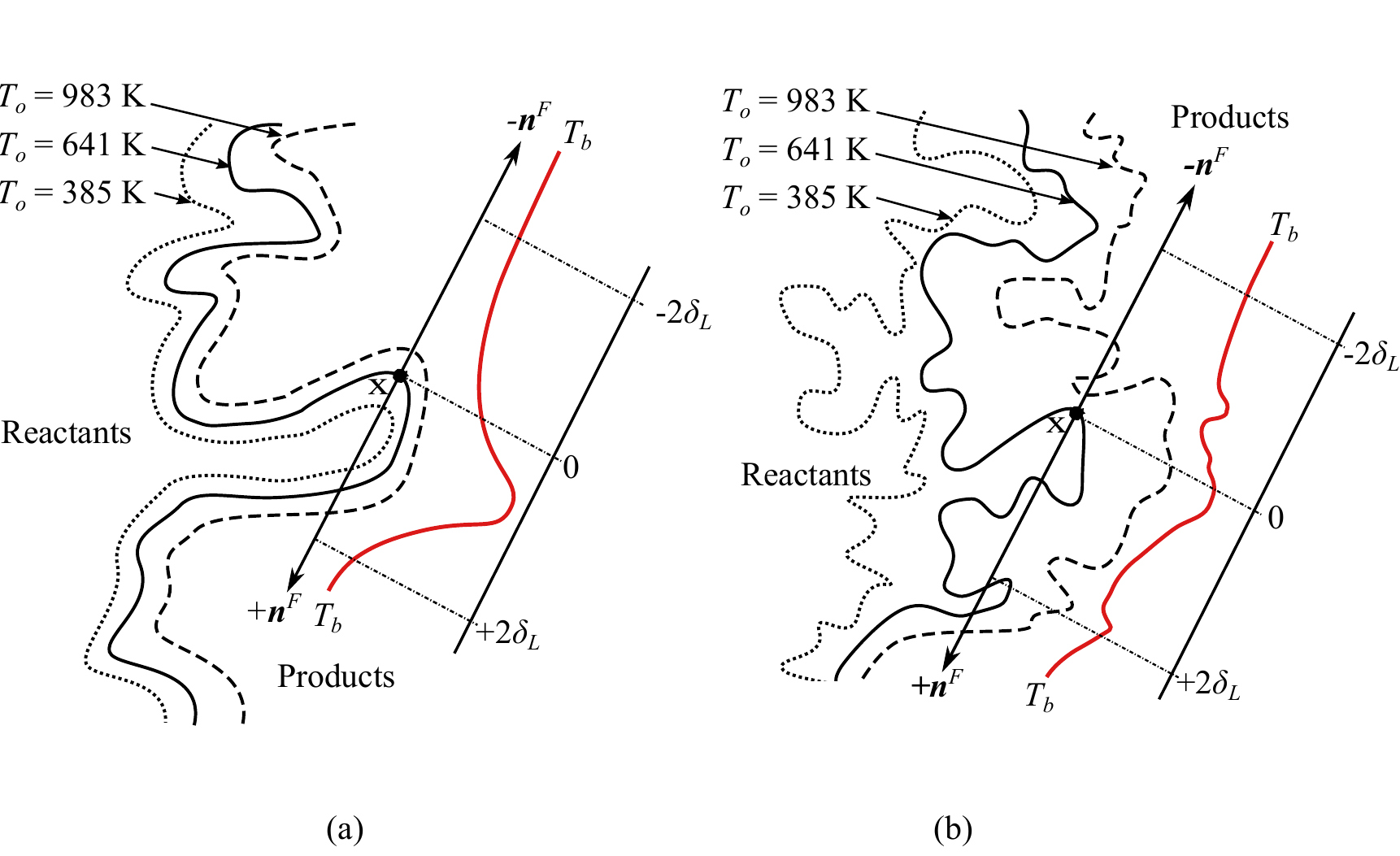}
\caption{Schematic representing flame-flame interaction that would be observed along the local surface normal, at the corresponding point $\bf{x}$ on the isotherm in its local neighbourhood for (a) F1 \& F3  (b) F2 \& F4  }
\label{Fig:Flame_Schematic}
\end{figure}

% The above questions are mostly addressed by the several sub-figures of Figs. \ref{Fig:Flame_Schematic}-\ref{Fig:F4_flame_structure}. In order to understand the curvature and flame shape effects, the local chemical and thermal flame structure is extracted from the three-dimensional DNS data. To generate the statistics of these local flame structures, first the flame locations for the four isotherms investigated are binned into eight different groups of $\widetilde{S_d}/S_L$. First group is $0 < \widetilde{S_d}/S_L \le 1$, second group is $1 < \widetilde{S_d}/S_L \le 2$ and so on. The final and eighth group is $\widetilde{S_d}/S_L > 7$. 

To this end, the DNS data were analyzed to extract the local chemical and thermal flame structures in the neighborhood of various points on the flame surface as follows. First, iso-scalar surfaces based on temperature ($T_0$ = constant) were identified from the three-dimensional DNS data. At any given point ($\textbf{x}$) on these isotherms, a one-dimensional flame structure is resolved along the direction normal to the front, $\boldsymbol{n}$ (Eq. \ref{eq:normal}). Fig.~\ref{Fig:Flame_Schematic} shows several isotherms representative of (a) moderately turbulent cases (F1 and F3) and (b) highly turbulent cases (F2 and F4). We define a line in 3D space using three points; one point is $\textbf{x}$ and other two points are at a distance of 2$\delta_L$ from $\textbf{x}$ in $+\boldsymbol{n}$ and $-\boldsymbol{n}$ directions evaluated at $\textbf{x}$. The line is then discretized using an appropriate number of points and the temperature and mass-fraction values from the 3D field are interpolated on these points. Thus, we obtain the local thermo-chemical structure over a distance of 4$\delta_L$. Fig.~\ref{Fig:Flame_Schematic} also shows the qualitative schematic of the temperature profiles obtained from this procedure. In this way, the flame shape could be identified as planar or cylindrical shapes depending on curvature at the points on the isotherm. For moderately turbulent cases (Fig.~\ref{Fig:Flame_Schematic}(a)), the global flame-flame interaction events dominate, and the local flame thickness remains nearly constant along the tangential directions of the surface, leading to near-parallel isotherms. The expected temperature profile shows a single valley point in the reactant mixture region. In contrast, for highly turbulent cases ((Fig.~\ref{Fig:Flame_Schematic}(b)), the non-parallel nature of the isotherms leads to multiple peaks and troughs in the interpolated temperature profile (shown by the red curve).

\begin{figure}[h!]
    %\begin{subfigure}[b]{1.0\textwidth}
    \centering\includegraphics[trim=1cm 1.5cm 1cm 1.5cm,clip,width=1.0\textwidth]{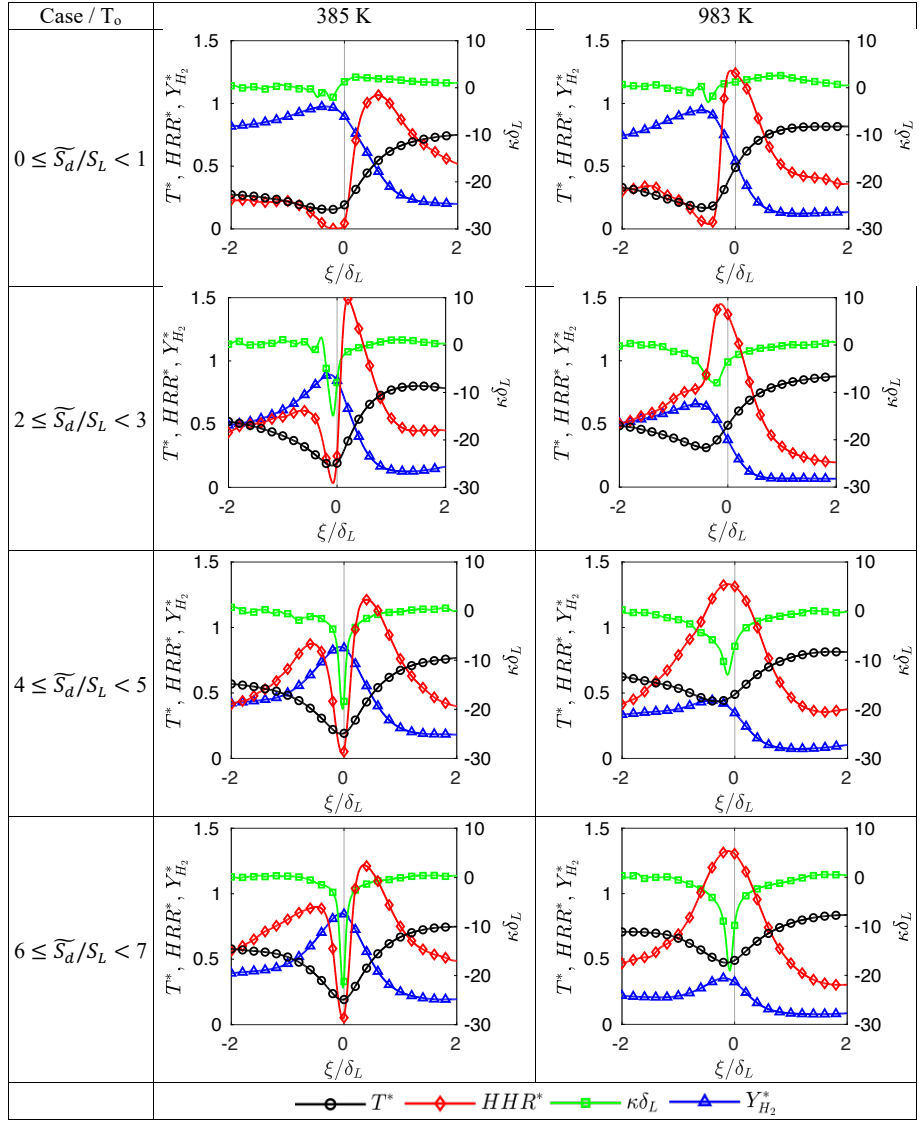}
    \caption{Local flame structure for Case F1 at $T_0 = 385 K$ and $T_0 = 983 K$}
    \label{Fig:F1_flame_structure}
    %\end{subfigure}
\end{figure}

\begin{figure}[h!]
   % \begin{subfigure}[b]{1.0\textwidth}
    \centering\includegraphics[trim=1cm 2cm 1cm 1cm,clip,width=1.0\textwidth]{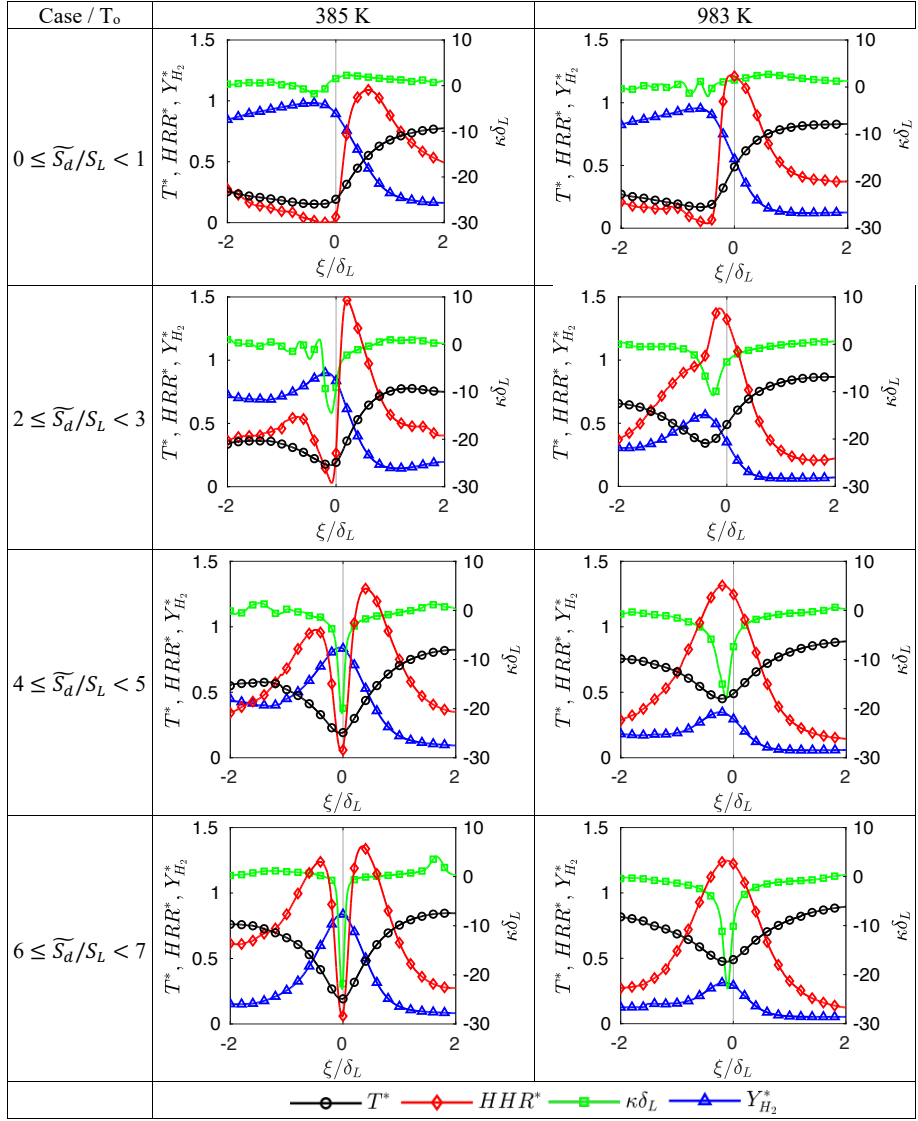}
    \caption{Local flame structure for Case F3 at $T_0 = 385 K$ and $T_0 = 983 K$}
    \label{Fig:F3_flame_structure}
  %  \end{subfigure}
\end{figure}

\begin{figure}[h!]
   % \begin{subfigure}[b]{1.0\textwidth}
    \centering\includegraphics[trim=1cm 3cm 1cm 0cm,clip,width=1.0\textwidth]{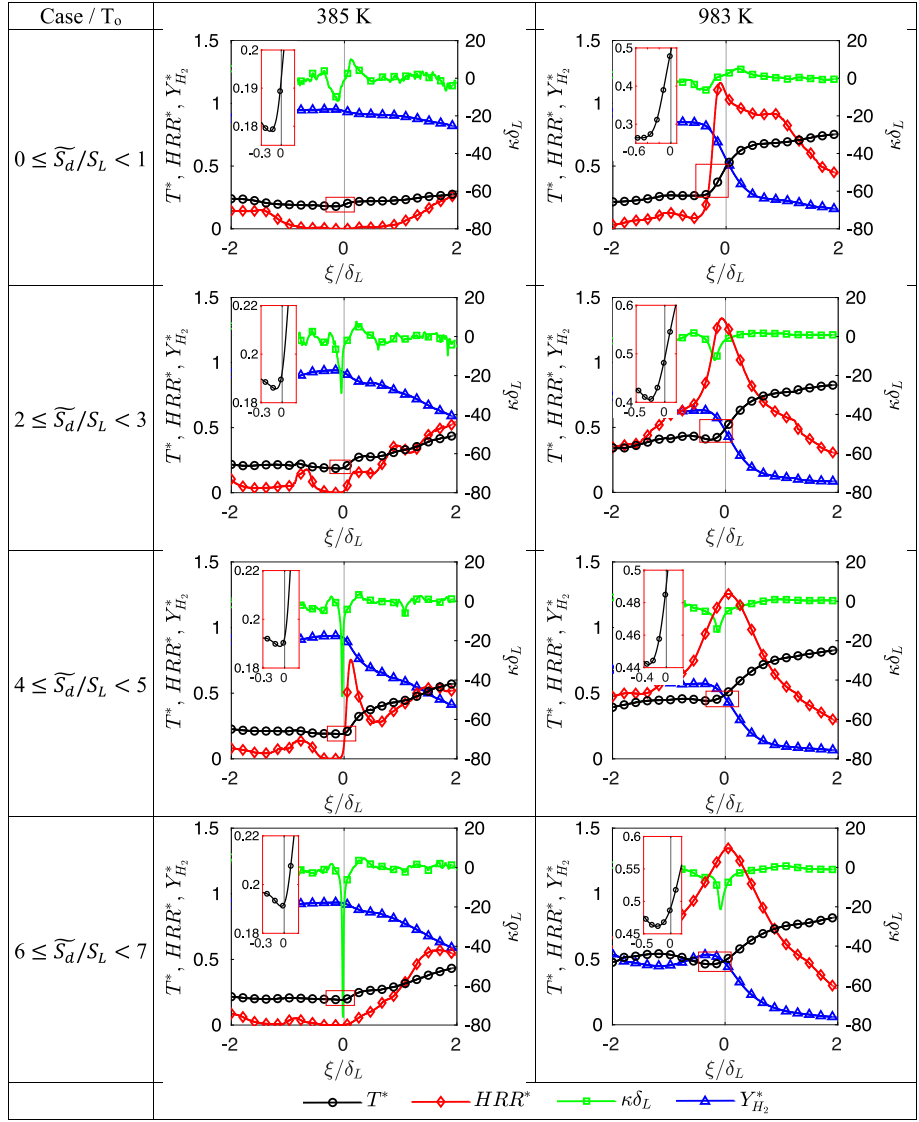}
    \caption{Local flame structure for Case F2 at $T_0 = 385 K$ and $T_0 = 983 K$}
    \label{Fig:F2_flame_structure}
    %\end{subfigure}
\end{figure}

\begin{figure}[h!]
\centering
     %\begin{subfigure}[b]{1.0\textwidth}
         \centering
         \includegraphics[trim=1cm 8cm 1cm 0cm,clip,width=1.0\textwidth]{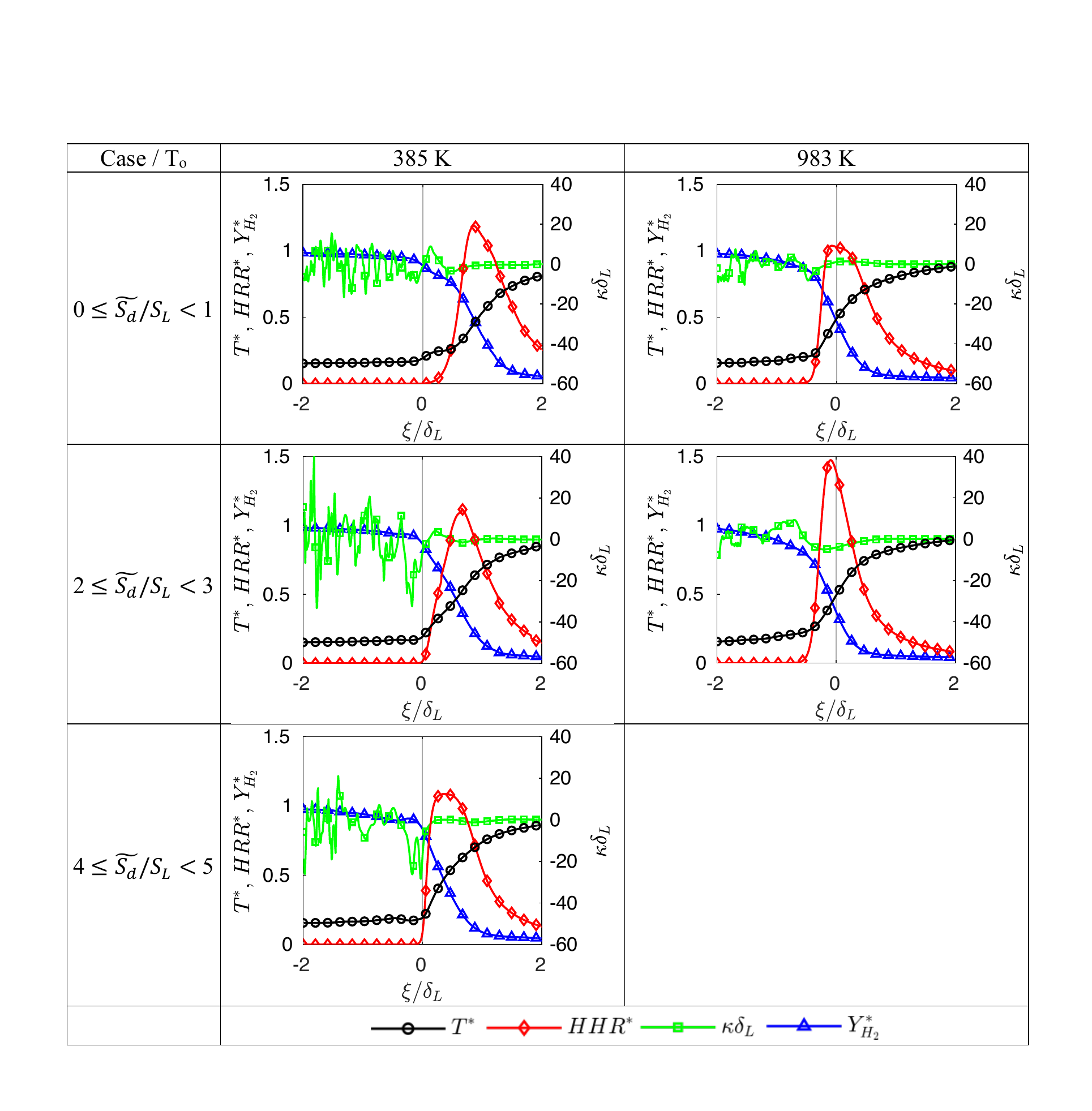}
         \caption{Local flame structure for Case F4 at $T_0 = 385 K$ and $T_0 = 983 K$}
         \label{Fig:F4_flame_structure}
     %\end{subfigure}
\end{figure}

To generate the statistics of the local flame structures, the points on the isotherms are grouped into eight different bins based on the increasing value of $\widetilde{S_d}/S_L$, with an increment of one, i.e. $0 \leq \widetilde{S_d}/S_L < 1$, $1 \leq \widetilde{S_d}/S_L < 2$ and so on, followed by averaging. The final and eighth group is $\widetilde{S_d}/S_L \geq 7$. Figs.~\ref{Fig:F1_flame_structure} - \ref{Fig:F4_flame_structure} presents the averaged structures for isotherms $T_0 = $ 385 K and 983 K for several binned groups of $\widetilde{S_d}/S_L$ for all the cases F1-F4. Here, $\xi$ denotes the local normal distance computed along $\boldsymbol{n}$. Thus, $\xi/\delta_L = 0$ denotes the point $\textbf{x}$ and $\xi/\delta_L \in [-2, 2]$, which is the abscissa in these figures. Point $\textbf{x}$ or $\xi/\delta_L = 0$ is marked by a faint vertical grey line in Figs.~\ref{Fig:F1_flame_structure} - \ref{Fig:F4_flame_structure}.

% \begin{figure}[h!]\ContinuedFloat
%     \begin{subfigure}[b]{1.0\textwidth}
%     \centering\includegraphics[trim=1cm 2cm 1cm 1.5cm,clip,width=1.0\textwidth]{figures/F1_local_flame_structure_b.pdf}
%     \caption{Local Flame Structure for $\widetilde{S_d}/S_L > 4$ for Case F1}
%     \label{Fig:F1_flame_structure_b}
%     \end{subfigure}
% \caption{Local flame structure for Case F1 at $T_0 = 385 K$ and $T_0 = 983 K$}
% \label{Fig:F1_flame_structure}
% \end{figure}

First, for the moderately turbulent conditions (Figs. \ref{Fig:F1_flame_structure} and  \ref{Fig:F3_flame_structure}) it is evident that an increase in $\widetilde{S_d}/S_L$ nearly correlates with a larger negative curvature, $\kappa\delta_L$, at the flame surface (point $\textbf{x}$). For positive $\kappa\delta_L$, $0 \leq \widetilde{S_d}/S_L < 1$. A large negative curvature implies an increased level of the upstream flame-flame interactions, as also evident by the V-shaped temperature profiles for larger $\widetilde{S_d}/S_L$ values. These V-shaped thermal profiles for Cases F1 and F3 show resemblance to the instantaneous thermal structures reported by Dave \& Chaudhuri \cite{dave2020} due to flame-flame interactions.
With the increase in level set temperature from $T_0$ = 385 K to 983 K, the minimum temperature increases and the maximum $Y_{H_2}$ decreases, implying that the two flames are closer together towards the mutual annihilation.

Figs.~\ref{Fig:F1_flame_structure} and  \ref{Fig:F3_flame_structure} also show the heat-release rate profiles (red curves) interpolated locally in addition to thermal and chemical structure. For a standard premixed H$_2$-air flame with $\phi$ = 0.7, the maximum heat release rate occurs at $T\sim$ 895 K. The fact that the heat release rate profiles show two distinct peaks for the 385 K isotherm and a merged single peak for the 983 K isotherm suggests that these data points represent a typical upstream flame-flame interaction without any possibility of autoignition of the upstream gases between the flames. Thus, $\widetilde{S_d}/S_L$ enhancement observed here is attributed to the interacting flame theory, not due to the acceleration by autoignition front, as reported in (\cite{sankaran2005,Chen2006,hawkes2006,krishman2018,gruber2021}) for higher reactant temperature and/or pressure conditions. Incidentally, the fact that the two heat release rate peaks already merged into a single peak for the 983 K isotherm data is attributed to the unique structure of the hydrogen-air flames where the heat release peak is located further upstream relative to that in hydrocarbon flames. Therefore, 
while Dave \& Chaudhuri \cite{dave2020} considered only ``weak" thermo-diffusive interactions for developing the interaction model, the distinct heat release structure in the hydrogen-air flames opens a possibility that interaction of reaction layers may play a significant role in enhancing $S_d$, in addition to thermo-diffusive effects. This point will be discussed in the later section.

% \begin{figure}[h!]\ContinuedFloat
%     \begin{subfigure}[b]{1.0\textwidth}
%     \centering\includegraphics[trim=1cm 2cm 1cm 2cm,clip,width=1.0\textwidth]{figures/F3_local_flame_structure_b.pdf}
%     \caption{Local Flame Structure for $\widetilde{S_d}/S_L > 4$ for Case F3}
%     \label{Fig:F2_flame_structure_b}
%     \end{subfigure}
% \caption{Local flame structure for Case F3 at $T_0 = 385 K$ and $T_0 = 983 K$}
% \label{Fig:F3_flame_structure}
% \end{figure}

Fig.~\ref{Fig:F2_flame_structure} show the flame structure for Case F2 with the largest $Ka$ and smallest size of the Kolmogorov eddies, exhibiting the most significant level of disruption in the preheat zone resulting in a large number of local collisions among the isotherms. As such, the thermal and chemical structures deviate significantly from those observed in moderate $Ka$ cases F1 and F3.
Consequently, neither the overall V-shaped temperature profile or the two distinct heat release rate peaks are seen for $T_0 =$ 385 K. This is because the small-scale fluctuations in the preheat zone smear out the gradients in the upstream region when averaged over the width of the flame thickness, implying that this cannot be described by the collision between entire flame structures. In this paper, the collision between complete flame structures will be referred to as ``global" collision as opposed to ``local" collision where isolated isotherms collide. Multiple peaks and small levels of fluctuations in the heat release rate and temperature are observed upstream of the flame.
These fluctuations in temperature are present over a relatively short distance ($\sim 0.5\delta_L$) in the upstream (a more detailed view is shown in the inset). Such a temperature profile implies that local flame-flame interactions associated with micro-mixing are prevalent in this condition. Therefore, while the large increase in $\widetilde{S_d}/S_L$ is still correlated with negative curvature, the statistically averaged flame structures are different from what was described by the flame interaction model.

Fig.~\ref{Fig:F4_flame_structure} shows the statistically averaged flame structure for Case F4. While $Ka$ = 721 is still large, $Re_t$ = 129 is much lower than that for Case F2, resulting in a much more suppressed level of the flame wrinkling \cite{song2020dns}. The lower $Re_t$ implies that the small scale eddies dissipate quickly and do not impart the same extent of the flame front wrinkling to promote the flame-flame collision as in Case F2. As a result, the thermal and chemical structures are similar to that of a planar laminar flame without a noticeable sign of interaction, while a small level of fluctuations are seen in the curvature of the upstream isotherms. 

% \begin{figure}[h!]\ContinuedFloat
%     \begin{subfigure}[b]{1.0\textwidth}
%     \centering\includegraphics[trim=1cm 3cm 1cm 1cm,clip,width=1.0\textwidth]{figures/F2_local_flame_structure_b.pdf}
%     \caption{Local Flame Structure for $\widetilde{S_d}/S_L > 4$ for Case F2}
%     \label{Fig:F3_flame_structure_b}
%     \end{subfigure}
% \caption{Local flame structure for Case F2 at $T_0 = 385 K$ and $T_0 = 983 K$}
% \label{Fig:F2_flame_structure}
% \end{figure}

%F3 lies in the thin reaction zone with $Ka\sim\mathcal{O}$(10) at low $Re_t$. Since we are again confined to the moderately turbulent premixed flames, the thermal and chemical profiles show resemblance to that of F1 (see Fig.~\ref{Fig:F3_flame_structure}). The heat release profile is also similar for all the level set temperature values (see Appendix). Similar to F1, the collisions among the flame surfaces are global. Consequently, the applicability of the interaction model is justified for F3.

% \begin{figure}[h!]\ContinuedFloat
%      \centering
%      \begin{subfigure}[b]{1.0\textwidth}
%          \centering
%          \includegraphics[trim=1cm 13cm 1cm 1.5cm,clip,width=1.0\textwidth]{figures/F4_local_flame_structure_b.pdf}
%          \caption{Local Flame Structure for $4 < \widetilde{S_d}/S_L < 6$ for Case F4}
%          \label{Fig:F4_flame_structure_b}
%      \end{subfigure}
% \caption{Local flame structure for Case F4 at $T_0 = 385 K$ and $T_0 = 983 K$}
% \label{Fig:F4_flame_structure}
% \end{figure}

\begin{figure}[h!]
\centering\includegraphics[trim=0.5cm 8cm 0.2cm 2cm,clip,width=1.0\textwidth]{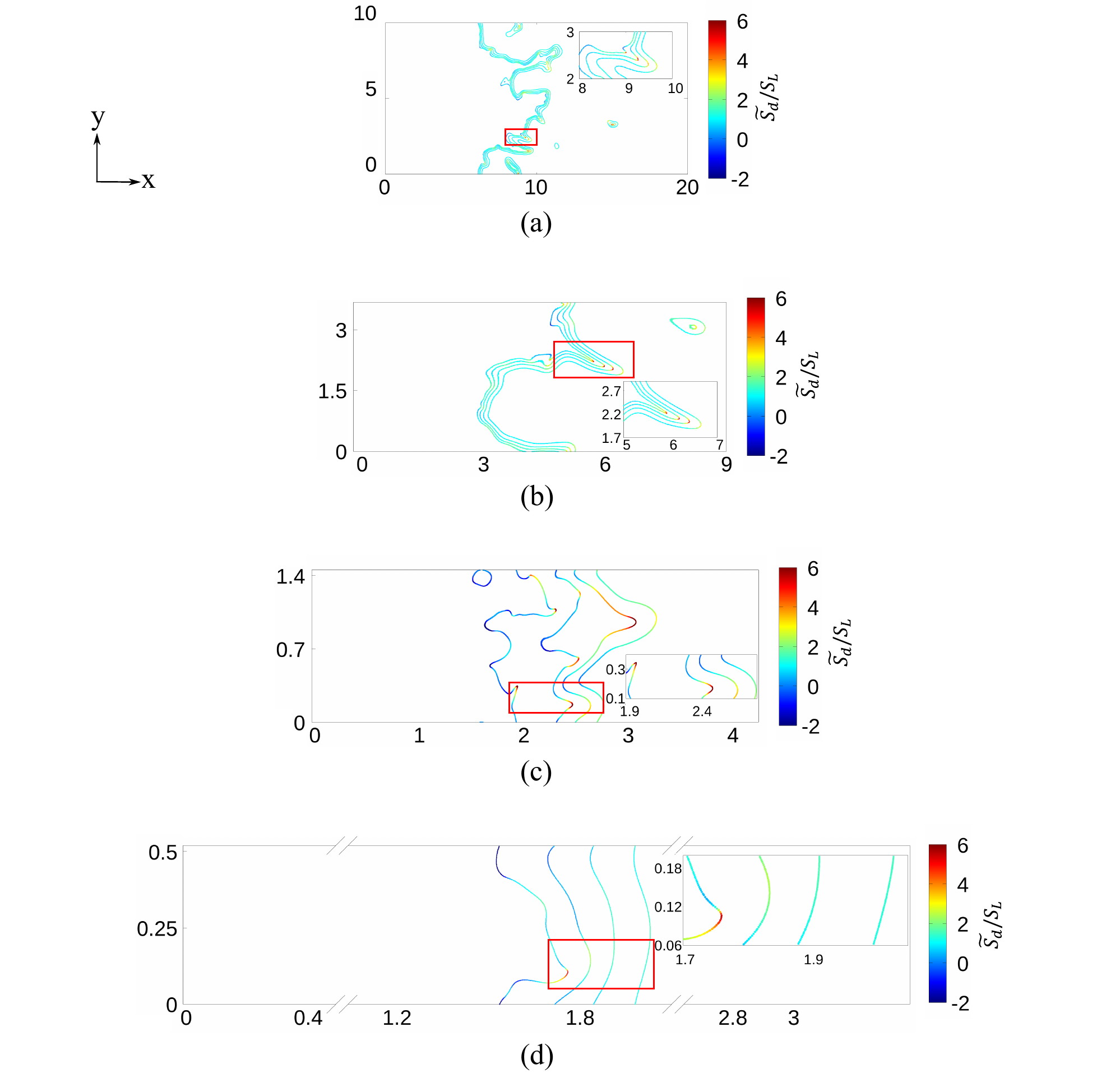}
\caption{Cross-sectional views of isotherms, $T_0 =$ 385, 641, 983 and 1325 K for cases (a) F1 (b) F3 (c) F2 (d) F4. The colorscale represents the $\widetilde{S_d}/S_L$}
\label{Fig:2D_Flame_sections}
\end{figure}

Fig.~\ref{Fig:2D_Flame_sections} shows cross-sectional views of isotherms ($T_0 =$ 385, 641, 983, 1325 K) with colors denoting the level of the  $\widetilde{S_d}/S_L$ enhancement. The regions of flame-flame interaction (both global and local) subjected to a large negative curvature are enclosed in red boxes with the close up view in insets. Large $\widetilde{S_d}/S_L$ values are observed in the event of flame-flame interactions, either global (Cases F1 (a) and F3 (b)) or local (Cases F2 (c) or F4 (d)). In particular,  Fig.~\ref{Fig:2D_Flame_sections}(d) (Case F4) reveals that the negative curvature becomes attenuated towards the downstream at higher temperature isotherms, at the given instant of interaction. Consequently, the $\widetilde{S_d}/S_L$ also decreases. % as the interaction prospects between the preheat zones decrease. 
Hence, for T = 983 K the maximum $\widetilde{S_d}/S_L$ observed is less than 3 compared to 6 observed for T = 385K (Figs.~\ref{Fig:SdSL_kappa} \& \ref{Fig:F4_flame_structure})

In summary, the statistical analysis of the isotherms showed that both global and local flame interactions in the highly curved regions lead to significantly enhanced local displacement speed. However, the inherent structure of hydrogen-air flames yielded an increased level of the heat release rate interaction located upstream, raising questions about the validity of the interacting flame theory. This issue is investigated in the following subsection by analyzing the individual terms contributing to the displacement speed.

\subsection{Budget analysis of ${S_d}$}
Here, a detailed term-by-term analysis is conducted based on Eq. \ref{eq: Sd_eq_DNS_T}. Only cases F1 and F2 are considered as representative moderate and high turbulence flame characteristics. Each of the terms in Eq. \ref{eq: Sd_eq_DNS_T} is normalized with the corresponding quantity at the same isotherm in the reference laminar premixed flame. The normalized terms, $\widetilde{T_1}$, $\widetilde{T_2}$ and $\widetilde{T_3}$ are written as: 

\begin{equation}\label{eq: T1_Normalized}
    \widetilde{T_1} = 
    \frac{|\nabla T|_L}{|\nabla T|}\frac{\nabla\cdot(\lambda^\prime\nabla T)}{(\nabla\cdot(\lambda^\prime\nabla T))_L}
\end{equation}
\begin{equation}\label{eq: T2_Normalized}
    \widetilde{T_2} = \frac{|\nabla T|_L}{|\nabla T|} \cdot \frac{\rho\nabla T \cdot \sum_{k}^{}  (\mathcal{D}_k C_{p,k} \nabla Y_k)}{(\rho\nabla T \cdot \sum_{k}^{}  (\mathcal{D}_k C_{p,k} \nabla Y_k))_L} 
\end{equation}
\begin{equation}\label{eq: T3_Normalized}
    \widetilde{T_3} = \frac{|\nabla T|_L}{|\nabla T|}\cdot \frac{\sum_{k}^{} h_k \dot{\omega}_k}{(\sum_{k}^{} h_k \dot{\omega}_k)_L}
\end{equation}
which represent the contributions from heat conduction, heat transfer by mass diffusion, and reaction, respectively. 

\begin{figure}[h!]
\centering\includegraphics[trim=2.0cm 4cm 2cm 2cm,clip,width=1.0\textwidth]{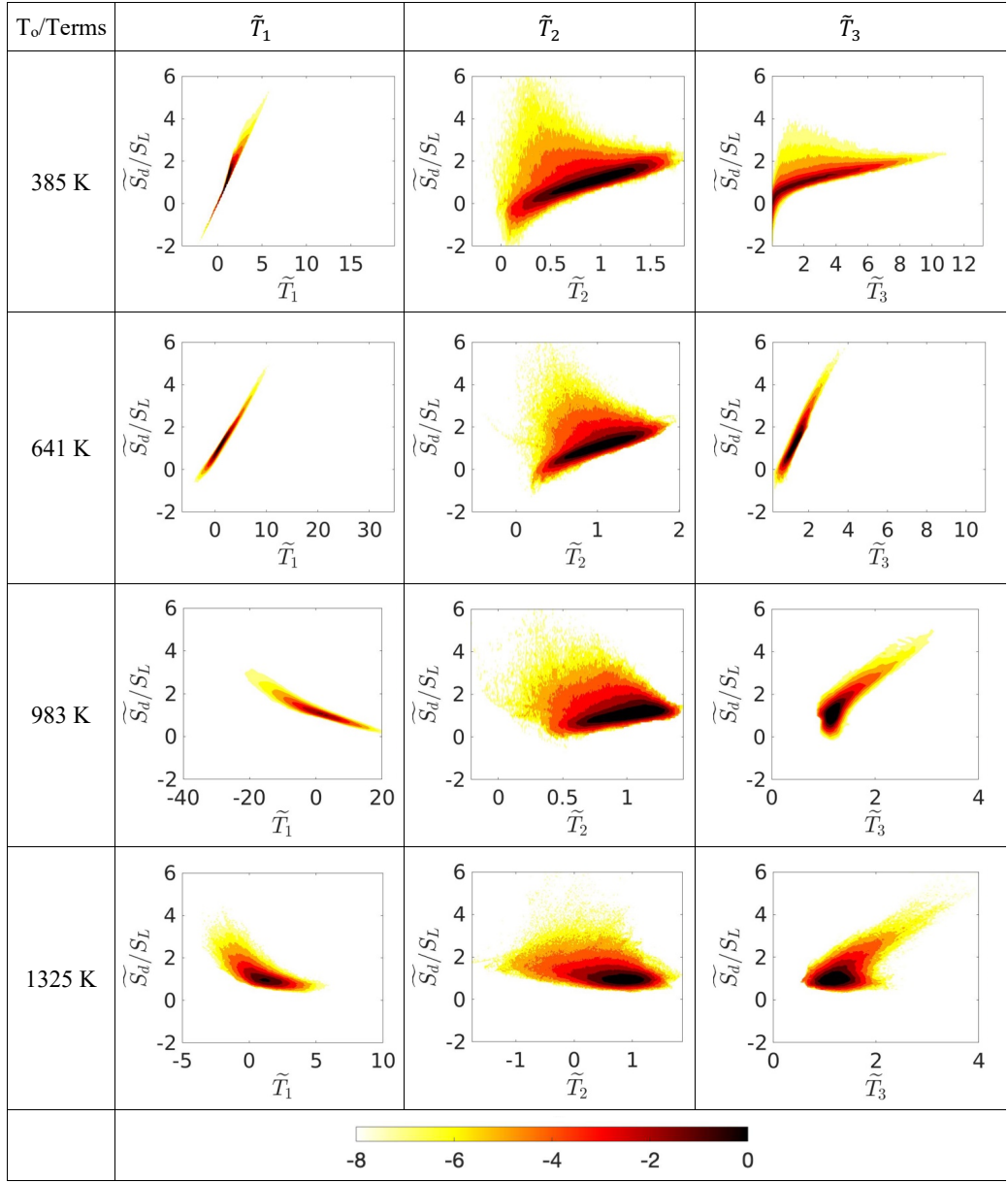}
\caption{JPDF of $\widetilde{S_d}/S_L$ with the terms $\widetilde{T_1}$, $\widetilde{T_2}$ and $\widetilde{T_3}$ for case F1. Colorscale represents natural logarithm of JPDF magnitudes.}
\label{Fig:F1_jpdf_term_ss}
\end{figure}

\begin{figure}[h!]
\centering\includegraphics[trim=2.0cm 4cm 2cm 2cm,clip,width=1.0\textwidth]{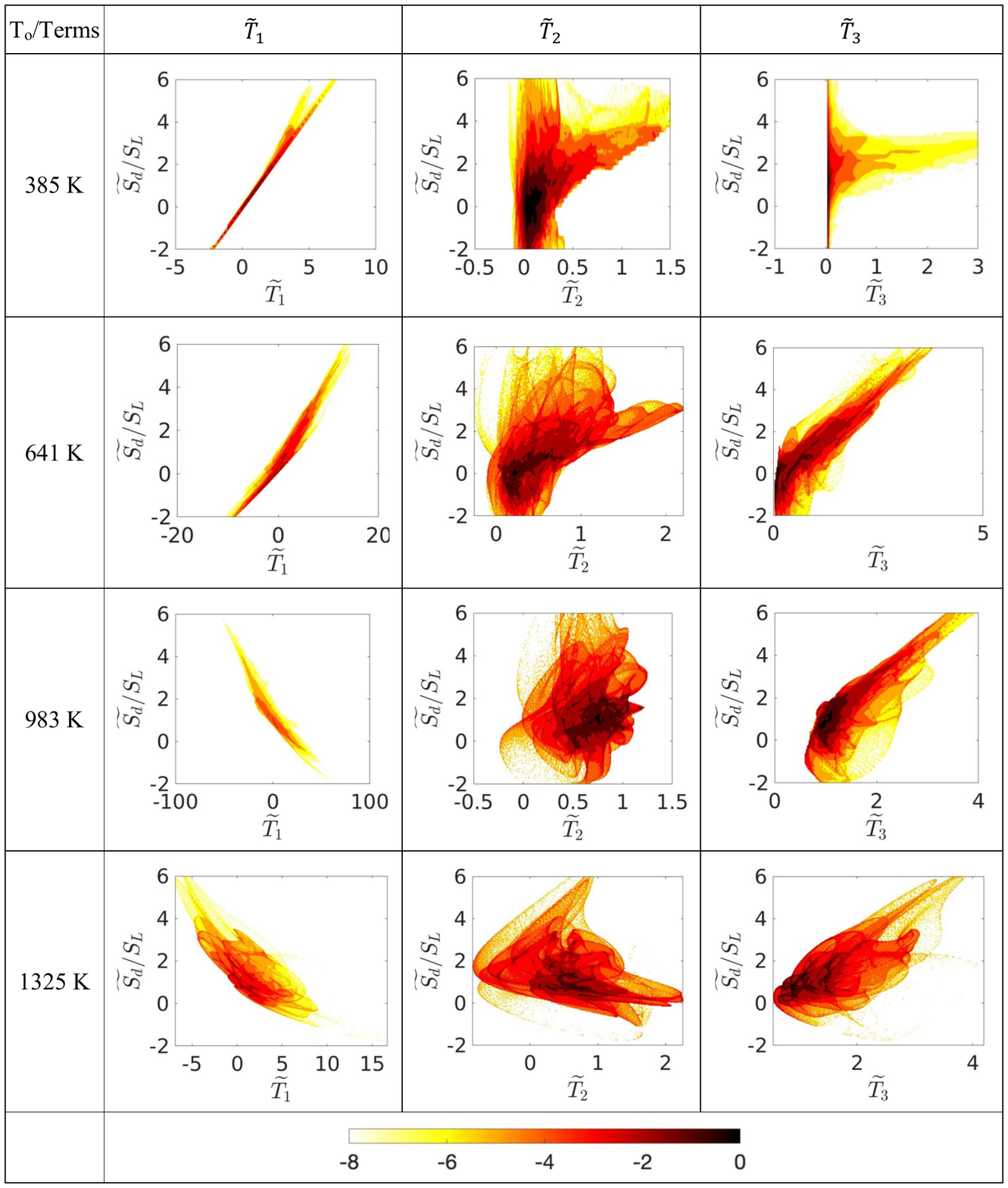}
\caption{JPDF of $\widetilde{S_d}/S_L$ with the  terms $\widetilde{T_1}$, $\widetilde{T_2}$ and $\widetilde{T_3}$ for case F2. Colorscale represents natural logarithm of JPDF magnitudes.}
\label{Fig:F2_jpdf_term_ss}
\end{figure}

The JPDF of $\widetilde{S_d}/S_L$ with each of the three terms are shown in Figs. \ref{Fig:F1_jpdf_term_ss} \& \ref{Fig:F2_jpdf_term_ss}. The colorscale at the bottom corresponds to natural logarithm of the JPDF values. 

% \begin{figure}[h!]
% \centering\includegraphics[trim=2.0cm 4cm 2cm 2cm,clip,width=1.0\textwidth]{figures/F3_JPDF_Term_ss.pdf}
% \caption{JPDF of $\widetilde{S_d}/S_L$ with the terms $\widetilde{T_1}$, $\widetilde{T_2}$ and $\widetilde{T_3}$ for case F3. Colorscale represents natural logarithm of JPDF magnitudes.}
% \label{Fig:F3_jpdf_term_ss}
% \end{figure}

% \begin{figure}[h!]
% \centering\includegraphics[trim=2.0cm 4cm 2cm 2cm,clip,width=1.0\textwidth]{figures/F4_JPDF_Term_ss.pdf}
% \caption{JPDF of $\widetilde{S_d}/S_L$ with the terms $\widetilde{T_1}$, $\widetilde{T_2}$ and $\widetilde{T_3}$ for case F4. Colorscale represents natural logarithm of JPDF magnitudes.}
% \label{Fig:F4_jpdf_term_ss}
% \end{figure}

Starting with $\widetilde{T_1}$, a strong positive correlation between $\widetilde{S_d}/S_L$ and $\widetilde{T_1}$ is observed for isotherms $T_0 =$ 385 K and 641 K, while the correlation turns negative for 983 K and 1325 K. The change in sign is associated with the maximum heat release rate that occurs at around 895 K. From Eq. \ref{eq: Sd_eq_DNS_T}, $T_1$ is written as,

\begin{equation}
    T_1 = \frac{\nabla \lambda^\prime \cdot \nabla T}{|\nabla T|}+ \frac{\lambda^\prime \nabla^2 T}{|\nabla T|}
\label{eq:t1}
\end{equation}\vspace{0.5cm}

% From \cite{dave2020}, the assumption of $\lambda^\prime \sim T$ is followed. 
As the upstream flame-flame interaction progresses, the temperature gradient decreases ($|\nabla T| \rightarrow $ 0) due to the increase in minimum temperature until the eventual annihilation. 
Using $\lambda^\prime \sim T^{0.7}$ and $\nabla T = -\boldsymbol{n}|\nabla T|$, we simplify the first term on the right hand side of Eq. \ref{eq:t1} as follows
%Thus, using $\nabla^2 T = \nabla\cdot(-\boldsymbol{n}|\nabla T|)$, Eq.\ref{eq:t1} is simplified to

\begin{equation}
    \begin{split}
     \frac{ \nabla\lambda^\prime\cdot\nabla T}{|\nabla T|} & =  - \nabla T^{0.7}\cdot\boldsymbol{n} \\
 %    & = -\Big[\frac{\partial}{\partial x}(T^{0.7})\hat{i}+\frac{\partial}{\partial y}(T^{0.7})\hat{j}+\frac{\partial}{\partial z}( T^{0.7})\hat{k}\Big]\cdot\boldsymbol{n} \\
%     \text{Applying chain rule}, \\
 %    & = -\Big[\frac{0.7}{T^{0.3}}\Big\{\frac{\partial}{\partial x}(T)\hat{i}+\frac{\partial}{\partial y}(T)\hat{j}+\frac{\partial}{\partial z}( T)\hat{k}\Big\} \Big]\cdot\boldsymbol{n} \\
%     & = -\frac{0.7}{T^{0.3}} \nabla T \cdot \boldsymbol{n} \\
%     & = -\frac{0.7}{T^{0.3}} (-\boldsymbol{n}|\nabla T|)\cdot \boldsymbol{n} \\
     & = \frac{0.7}{T^{0.3}}|\nabla T|
     \nonumber
    \end{split}
\end{equation}
Thus given the assumption that during interaction $|\nabla T| \rightarrow$ 0, $T_1$ reduces to,

\begin{equation}
    T_1 = - \frac{ \lambda^\prime}{|\nabla T|} \nabla \cdot (\boldsymbol{n}|\nabla T|)
\end{equation}
With further simplification we get,
\begin{equation}
    T_1 = - \lambda^\prime\kappa - \frac{ \lambda^\prime}{|\nabla T|} \boldsymbol{n} \cdot \nabla (|\nabla T|)
\end{equation}
For a planar laminar premixed flame, $\kappa =$ 0  we get $(T_1)_L$ as,

\begin{equation}
    (T_1)_L =  - \frac{ \lambda^\prime}{|\nabla T|_L} \boldsymbol{n} \cdot \nabla (|\nabla T|_L)
\end{equation}
Thus Eq. \ref{eq: T1_Normalized} for $\widetilde{T_1}$ can be written as,
\begin{equation}
    \widetilde{T_1} = \frac{- \kappa - \frac{ 1}{|\nabla T|} \boldsymbol{n} \cdot \nabla (|\nabla T|)}{- \frac{ 1}{|\nabla T|_L} \boldsymbol{n} \cdot \nabla (|\nabla T|_L)}
\end{equation}
With $\boldsymbol{n}$ as the local unit normal vector, substituting  $|\nabla T| = \partial T/ \partial n$ and $\boldsymbol{n} \cdot \nabla = \partial / \partial n$ in above equation the final form of $\widetilde{T_1}$ becomes,

\begin{equation}
    \widetilde{T_1} = \frac{- \kappa - \frac{\partial^2 T/ \partial n^2}{\partial T/ \partial n}}{- (\frac{ \partial^2 T/ \partial n^2}{\partial T/ \partial n})_L}
\label{eq:t1final}
\end{equation}

Eq. \ref{eq:t1final} clearly shows the sign change of the term is due to $\partial^2 T/ \partial n^2$ which changes from positive to negative value beyond the maximum heat release point. In the meantime, $\partial T/ \partial n$ remains negative. Thus the correlation between $\widetilde{S_d}/S_L$ and $\widetilde{T_1}$ turns negative from 641 K to 983 K. More importantly, the numerator $T_1$ retains its positive correlation due to the fact that it is dominated by large curvature $\kappa$, overriding the effect of $\frac{ \partial^2 T/ \partial n^2}{\partial T/ \partial n}$.

Term $T_2$ is found to be smaller in magnitude compared to the terms $T_1$ and $T_3$ by one to two orders of magnitude. Furthermore, $\widetilde{T_2}$ shows large scatters and a very weak correlation with the $\widetilde{S_d}/S_L$. 

Lastly, $\widetilde{T_3}$ is related to the heat release rate. While the correlations are weak for the isotherms $T_0 =$ 385 K and 1325 K, a noticeable positive correlation is found for $T_0 =$ 641 K and 983 K, which are two isotherms in the neighborhood of the maximum heat release rate. Evidently, the contribution of $\widetilde{T_3}$ to the $\widetilde{S_d}/S_L$ enhancement occurs only for the isotherms in the near vicinity of maximum heat release rate; otherwise $\widetilde{S_d}$ remains unaffected. The nonuniform contribution of $\widetilde{T_3}$ to $\widetilde{S_d}$ throughout the flame makes the heat release term extremely important. 

\begin{figure}[h!]
\centering\includegraphics[trim=2.0cm 10cm 2cm 3cm,clip,width=1.0\textwidth]{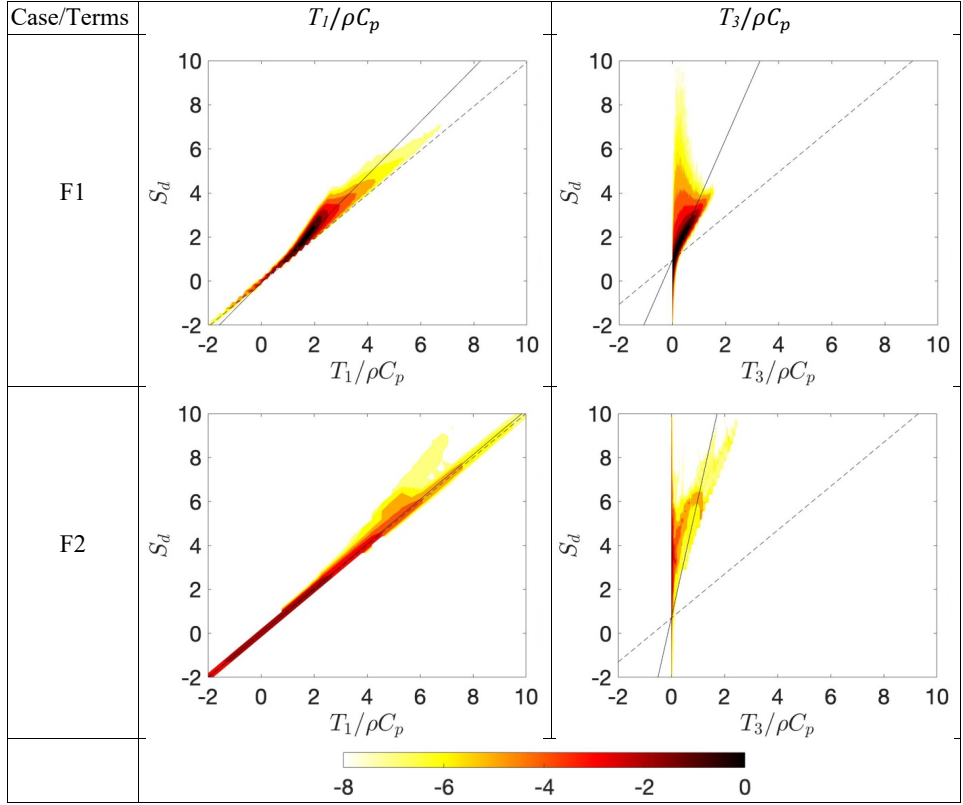}
\caption{Joint probability distribution function of $S_d$ with $T_1/\rho C_p$ and $T_3/\rho C_p$ for case F1 and F2 at $T_0 =$ 385 K. The  solid black line presents the linear curve of best fit. The dashed line has unit slope and intercept equal to the line of best fit. Colorscale represents natural logarithm of JPDF magnitudes.}
\label{Fig:JPDF_385K_F1_F2_T1_T3}
\end{figure}

% \begin{table}[h!]
% \footnotesize
%  \begin{center}
%  \begin{tabular}{ | p{3.2cm} |  C{2cm} | C{2cm} | C{2cm} | C{2cm} |} 
%  \hline
%  Cases/Terms & \multicolumn{2}{c|}{$T_1/\rho C_p$} & \multicolumn{2}{c|}{$T_3/\rho C_p$} \\ 
%  \cline{2-5}
%  & $P_1$ & $P_2$ & $P_1$ & $P_2$ \\
%  \hline
%  \hline
%  F1 & 1.2190 & -0.0701 & 2.7400 & 0.9425 \\
%  F2 & 1.0150 & 0.0079 & 5.3900 & 0.6943 \\
%  \hline
% \end{tabular}
% \caption{\label{tab:2} The slope ($P_1$) and intercept ($P_2$) for the line of best fit for F1 and F2 at $T_0$ = 385 K.}
% \end{center}
% \end{table}

\begin{figure}[h!]
\centering\includegraphics[trim=2.0cm 10cm 2cm 3cm,clip,width=1.0\textwidth]{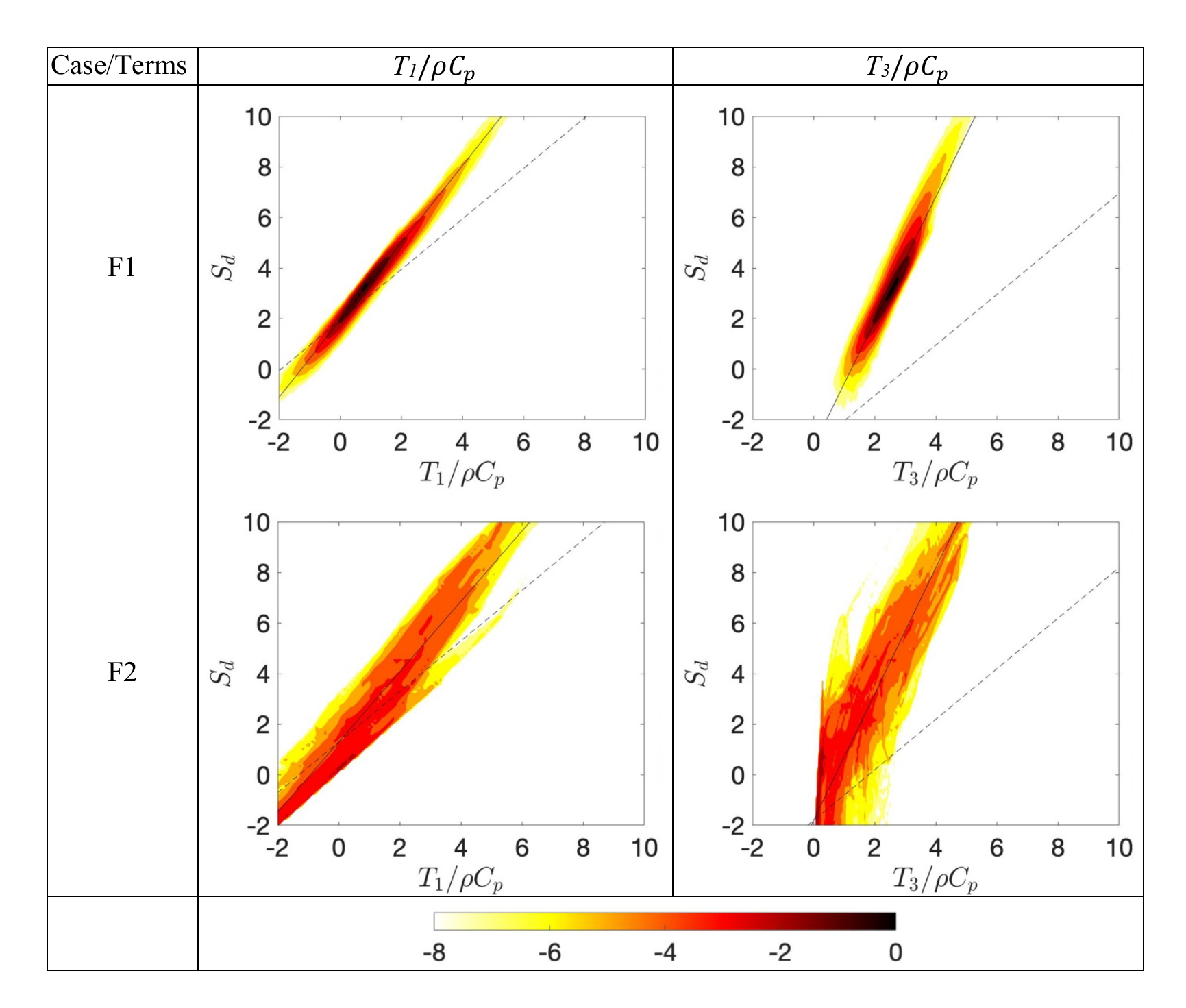}
\caption{Joint probability distribution function of $S_d$ with $T_1/\rho C_p$ and $T_3/\rho C_p$ for case F1 and F2 at $T_0 =$ 641 K. The  solid black line presents the linear curve of best fit. The dashed line has unit slope and intercept equal to the line of best fit.Colorscale represents natural logarithm of JPDF magnitudes.}
\label{Fig:JPDF_641K_F1_F2_T1_T3}
\end{figure}

% \begin{table}[h!]
% \footnotesize
%  \begin{center}
%  \begin{tabular}{ | p{3.2cm} | C{2cm} | C{2cm} | C{2cm} | C{2cm} |} 
%  \hline
%  \multirow{2}{3.2cm}{Cases/Terms} & \multicolumn{2}{c|}{$T_1/\rho C_p$} & \multicolumn{2}{c|}{$T_3/\rho C_p$} \\ 
%  \cline{2-5}
%  & $P_1$ & $P_2$ & $P_1$ & $P_2$ \\
%  \hline
%  \hline
%  F1 & 1.5270 & 1.9370 & 2.4640 & -3.0470 \\
%  F2 & 1.3920 & 1.2950 & 2.4980 & -1.8060 \\
%  \hline
% \end{tabular}
% \caption{\label{tab:3}The slope ($P_1$) and intercept ($P_2$) for the line of best fit for F1 and F2 at $T_0$ = 641 K.}
% \end{center}
% \end{table}

\begin{table}[h!]
\footnotesize
 \begin{center}
 \begin{tabular}{ | p{1.8cm} |  C{1cm} | C{1cm} | C{1cm} | C{1cm} | C{1cm} | C{1cm} | C{1cm} | C{1cm} |} 
 \hline
 & \multicolumn{4}{c|}{$T_0$ = 385 K} & \multicolumn{4}{c|}{$T_0$ = 641 K} \\ 
  \cline{2-9}
 Cases/Terms & \multicolumn{2}{c|}{$T_1/\rho C_p$} & \multicolumn{2}{c|}{$T_3/\rho C_p$} & \multicolumn{2}{c|}{$T_1/\rho C_p$} & \multicolumn{2}{c|}{$T_3/\rho C_p$}\\ 
 \cline{2-9}
 & $P_1$ & $P_2$ & $P_1$ & $P_2$ & $P_1$ & $P_2$ & $P_1$ & $P_2$\\
 \hline
 \hline
 F1 & 1.219 & -0.070 & 2.740 & 0.9425 & 1.527 & 1.937 & 2.464 & -3.047 \\
 F2 & 1.015 & 0.008 & 5.390 & 0.6943 & 1.392 & 1.295 & 2.498 & -1.806\\
 \hline
\end{tabular}
\caption{\label{tab:2} The slope ($P_1$) and intercept ($P_2$) for the line of best fit for F1 and F2 at $T_0$ = 385 K \& $T_0$ = 641 K.}
\end{center}
\vspace{-0.9cm}
\end{table}

To assess the role of heat release rate on $S_d$, Figs.~\ref{Fig:JPDF_385K_F1_F2_T1_T3} and \ref{Fig:JPDF_641K_F1_F2_T1_T3} present the JPDF of $S_d$ with $T_1/\rho C_p$ and $T_3/\rho C_p$ at $T_0$ = 385 K and 641 K, respectively. A line fit of the data is shown along with a line of unity slope. The term which contributes more to $S_d$ should have a slope closer to unity. Both lines have the same intercept. The slope ($P_1$) and intercept ($P_2$) values are tabulated in Table~\ref{tab:2}. The focus is to verify whether heat-release rate becomes an increasingly dominant term in controlling enhancement of $S_d$ in flames with high $Ka$. If so, the next question is whether this happens only close to the reaction zones. To address this, the magnitudes of $T_3$ relative to $T_1$ and its dependence on turbulence intensity are examined.

Fig.~\ref{Fig:JPDF_385K_F1_F2_T1_T3} for $T_0$ = 385 K shows the behavior far from the reaction zone. From F1 to F2, the increase in $S_d$ is due to larger values of $T_1$ and $T_3$, but $T_1$ continues to dominate over $T_3$. Thus, for isotherms closer to fresh reactants (that are also far from reaction zones) flame-flame interactions are predominantly governed by heat conduction. 
%$T_1$ shows a good correlation with $S_d$ compared to $T_3$ for both F1 and F2. 
%It should be noted that an increase in turbulence intensity from F1 to F2 leads to the strengthening of correlation between $T_1$ and $S_d$ and weakening of that between $T_3$ and $S_d$. 
Moving closer to the region of maximum heat release, i.e., for $T_0 =$ 641 K (Fig.~\ref{Fig:JPDF_641K_F1_F2_T1_T3}), however, the heat release rate term $T_3$ becomes more important. Moreover, $T_3$ takes larger values when $S_d$ is large, indicating that it is increasingly crucial in governing large enhancement of $S_d$ during flame-flame interactions. However, contribution by $T_3$ is still relatively small in magnitude compared to $T_1$, indicating that even at larger $Ka$ though $T_3$ becomes increasingly important, the overall interaction dynamics is still largely dependent on the diffusion ($T_1$) term. This increasing influence of heat-release rate term is pronounced for isotherms closer to the reaction zones and it rapidly diminishes away from it.

Therefore, the budget analysis showed that the density averaged displacement flame speed is predominantly dependent on $T_1$, i.e., large negative curvature ($\kappa \leq - 1/\delta_L$) of the isotherms as a result of internal flame-flame interactions for high $Re_t$ and $Ka$ premixed flames. This suggests that the interacting flame theory remains valid for highly turbulent premixed flames. The close agreement of the model and the DNS data shown in Fig. \ref{Fig:SdSL_kappa} can be attributed to physical reasons.

\section{Conclusions}

Three dimensional DNS datasets of lean H$_2$-air premixed flame with a wide range of $Re_t$ and $Ka$ from a previous study by \citet{song2020dns} were investigated with the purpose of assessing the validity of the interacting flame theory in high $Ka$ conditions. A detailed reaction mechanism involving 9 species and 23 reactions by \citet{burke2012comprehensive} was used. The interaction model proposed by \citet{dave2020} explained the enhancement of density-weighted flame displacement speed ($\widetilde{S_d}$) for negatively curved flame surfaces resulting from flame-flame interaction, for moderately turbulent flames ($Ka \sim\mathcal{O}$(10)). The present study showed that the model successfully captures the trends of $S_d$ with large negative curvature $\kappa$ for highly turbulent flames (up to $Ka \sim\mathcal{O}$(1000)).

The investigation of local thermal and chemical flame structures on the flame surface based over a range of $\widetilde{S_d}/S_L$ revealed the nature of flame-flame interactions for different cases. For moderately turbulent cases (F1 and F3), the complete flame structures collide, whereas F2 with $Ka\sim\mathcal{O}$(1000) showed local interaction of the individual flame surfaces - both leading to enhancement of $\widetilde{S_d}$. Unlike the first three cases, the local flame structures for F4 resembled that of a laminar flame in the absence of appreciable wrinkling due to turbulent dissipation. The flame structures also revealed the difference in heat release profile depending on the value of the level set temperature ($T_0$) due to the early interaction of the reaction layers for H$_2$-air flame. 

Analysis of individual terms of the local flame displacement speed (Eq.~\ref{eq: Sd_eq_DNS_T}) showed a strong correlation of the heat flux term with $S_d$ for all $T_0$. On the other hand, the heat release term contributed strongly to $S_d$ enhancement near the reaction zone and weakly when far from it. Even though the heat release term becomes increasingly important near the region of maximum heat release, the diffusion term still dominates with higher values contributing to large $S_d$ at large $-\kappa$ for all $Ka$. The interacting flame theory considered the diffusion term alone, under the assumption of the isotherm being sufficiently far from the reaction zone. Thus the proposed interaction model proved appropriate for premixed flames under intense turbulence as well.

\section{Acknowledgement}

This research was enabled in part by support provided by the Natural Sciences and Engineering Research Council of Canada through a Discovery Grant, Heuckroth Distinguished Faculty Award in Aerospace Engineering from UTIAS, and support from Compute Canada. In addition, computational resources were provided by KAUST Supercomputing Laboratory (KSL), alongside support from KAUST.



%% References



%% New version of the num-names style
\bibliographystyle{elsarticle-num-names}
\bibliography{sample.bib}






% \end{thebibliography}

\end{document}